\documentclass[epj]{svjour}
\usepackage{latexsym}
\usepackage{graphics}
\usepackage{amssymb}
\begin{document}
\title{Solid phase epitaxy amorphous silicon regrowth : some insight from empirical molecular dynamics simulation}
\subtitle{Insight from molecular dynamics simulation on amorphous silicon re-growth}
\titlerunning{Some insight from molecular dynamics simulation on amorphous silicon re-growth}
\author{C. Krzeminski \inst{1} \and E. Lampin \inst{1}% etc
}                     % Do not remove
\institute{Institut d'Electronique, de Micro\'electronique et de Nanotechnologies, D\'epartement ISEN, Avenue Poincar\'e, Cit\'e Scientifique, BP 60069, 59652 Villeneuve d'Ascq Cedex, France}

\mail{christophe.krzeminski@isen.fr}         % Insert a name or remove this line

\date{Received: date / Revised version: date}
\abstract{The modelling of interface migration and the associated diffusion mechanisms  at the nanoscale level is a challenging issue. For many technological applications ranging from nanoelectronic devices to solar cells, more knowledge of the mechanisms governing the  migration of the silicon amorphous/crystalline interface and dopant diffusion during solid phase epitaxy is needed. In this work, silicon recrystallisation in the framework of solid phase epitaxy and the influence on orientation effects  have been investigated at the atomic level using empirical molecular dynamics simulations. The morphology and the migration process of the interface has been observed to be  highly dependent on the original inter-facial atomic structure.  The  $[100]$ interface migration is a quasi-planar ideal process whereas the  cases  $[110]$ and $[111]$ are much more complex with a more diffuse interface. For $[110]$, the interface migration corresponds to the formation and dissolution of nanofacets whereas for $[111]$ a defective based bilayer reordering  is the dominant re-growth process. The study of the  interface velocity migration in the ideal case of defect free re-growth  reveals no difference between $[100]$ and $[110]$ and  a decrease by a mean factor of 1.43  for the case  $[111]$. Finally, the influence of boron atoms in the amorphous part on the interface migration velocity is also investigated in the case of $[100]$ orientation.} 
%end of abstract
%
\maketitle
\newpage
\section{Introduction}
\label{intro}

Advanced technological research work dedicated to silicon materials process engineering  needs to control the silicon recrystallisation process at the nanometer scale for nanoelectronic \cite{Duffy07,Saenger07} or photo-voltaic devices \cite{Agaiby,Huet09}. The silicon film re-growth occurs during annealing when amorphous regions or layers are created by  parasitic phenomena like in vapor-liquid-solid (VLS) grown silicon nanowires or intentionally, to limit  diffusion effects like in advanced  junction depth processing \cite{Duffy04}. However, it is really difficult to reach a deeper understanding of  the many effects taking place during regrowth. For example, the rich interplay between the nucleation rate, the interface roughening, the defects generation or dissolution and the dopant diffusion  is difficult to characterise by using only experimental fabrication and characterisation techniques.

From the theoretical point of view, there is a  debate  on the general physical description of crystal growth  \cite{Chernov04}. The concept of a localised interface approach where  the nucleation and growth process take place was first introduced by Burton, Cabrera and Frank \cite{BCF51}. On the other hand, the concept of a diffuse interface where an ordering flux is associated to the interface migration is also proposed. This flux is assumed to be  generated by the counterbalance  between the atomic forces interaction induced by the short-range  order  and the thermal motion \cite{Chernov04}. In this framework,  solid phase regrowth is often described as an thermal activated process shown  by the following equation :

\begin{equation}
\displaystyle
v_{SPE}(T)=v_{0}\exp \big [ -\frac{E_{a}}{kT} \big ]
\end{equation}

where T is the interface temperature, v$_{0}$ is the exponential prefactor and E$_{a}$ is the activation energy. For silicon phase epitaxy, the debate focused on the  nature of defects (dangling bonds, kink sites,  vacancies, interstitials) on their location (crystal, interface, amorphous part) and on their influence on the limited growth rate \cite{Aziz91}. The influence of interface orientation, stress or dopant is also under debate. Atomistic process modelling \cite{Pelaz09}  in the framework of molecular dynamics is expected  to help to clarify the physical picture. This method could probably  provide more insight into the re-growth mechanisms taking place at the amorphous/crystalline interface during thermal annealing. Dopant incorporation  and the associated diffusion effects during re-growth are much more complex to simulate compared to the bulk case \cite{Cuny06}. Moreover, information technology achieved significant progress and enabled  the simulation of relatively large atomic configurations  that could match  experimental systems.\\

Preliminary simulation studies of the silicon  solid phase epitaxy related to junction processing  by molecular dynamics were performed  by Mootoka {\it  et al.} \cite{Motooka00}. But many open questions remain on solid phase epitaxy like the substrate orientation influence  despite  clear technological interest. More knowledge is available in the case of melt-interfaces re-growth \cite{Grabow89}.  This problem has also been investigated  using a finite elements based method or lattice Kinetic Monte Carlo (KMC) \cite{Bragado09} but numerous challenges are faced to take into account the complex reordering processes at the atomic level. The main objective of this paper is to present the work performed in order to investigate the influence of  orientation and dopant effects on  the amorphous silicon re-growth mechanisms. The first section briefly describes  the  computational details  and  the method used in this work. Some complementary tests related to the influence of the initial properties of the disordered layer are also provided.  In a second part, most important  results concerning the influence of the interface orientation on  recrystallisation mechanisms are presented and discussed. Preliminary results on the influence of Boron atoms on the interface migration velocity are also provided. Finally, some important conclusions are drawn in the last section.

\section{Methods}

\subsection{Computational details}

\begin{figure}[tbp]
\centering

\resizebox{0.75\columnwidth}{!}{
\includegraphics{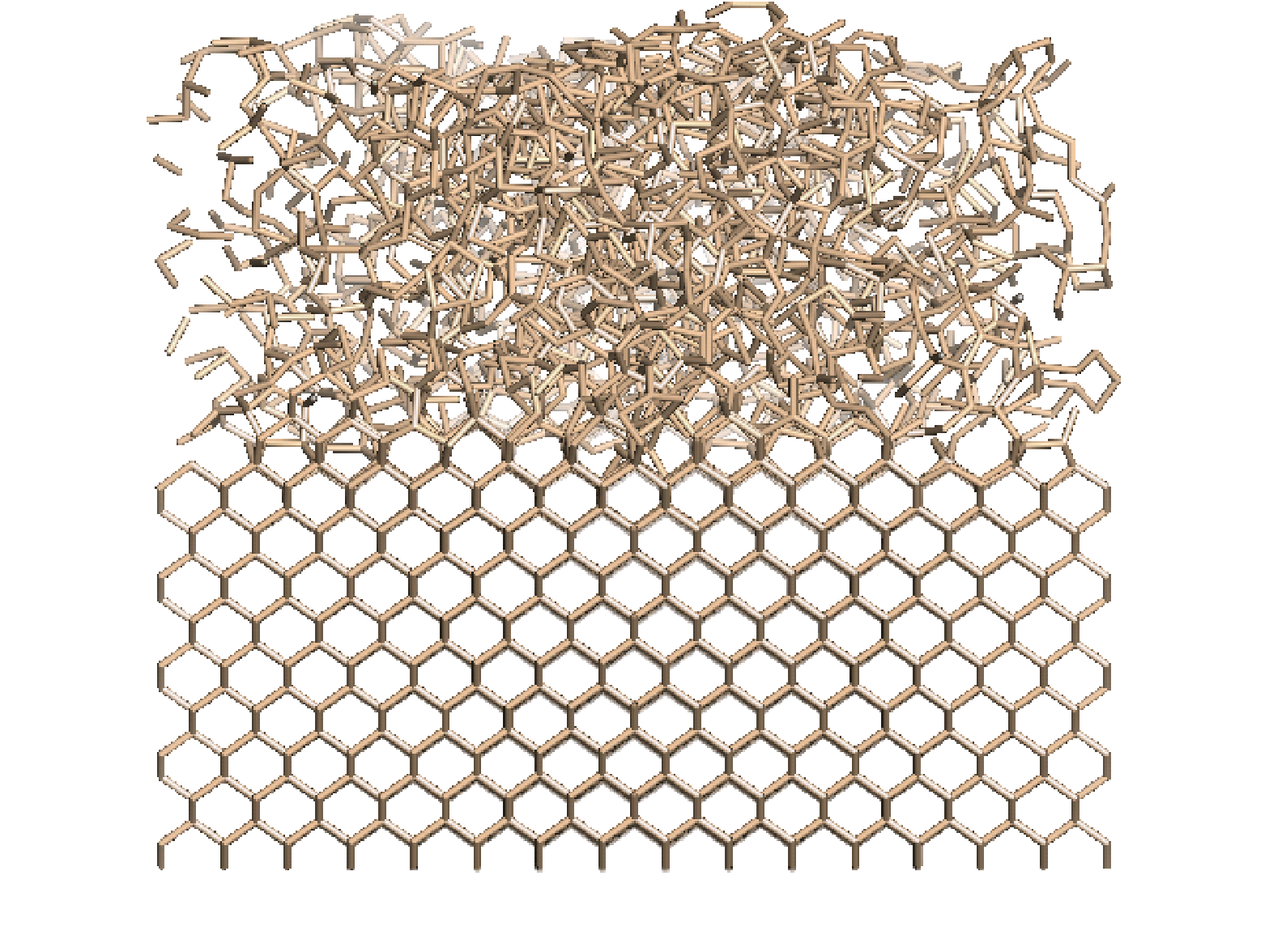}}
\resizebox{0.75\columnwidth}{!}{
\includegraphics{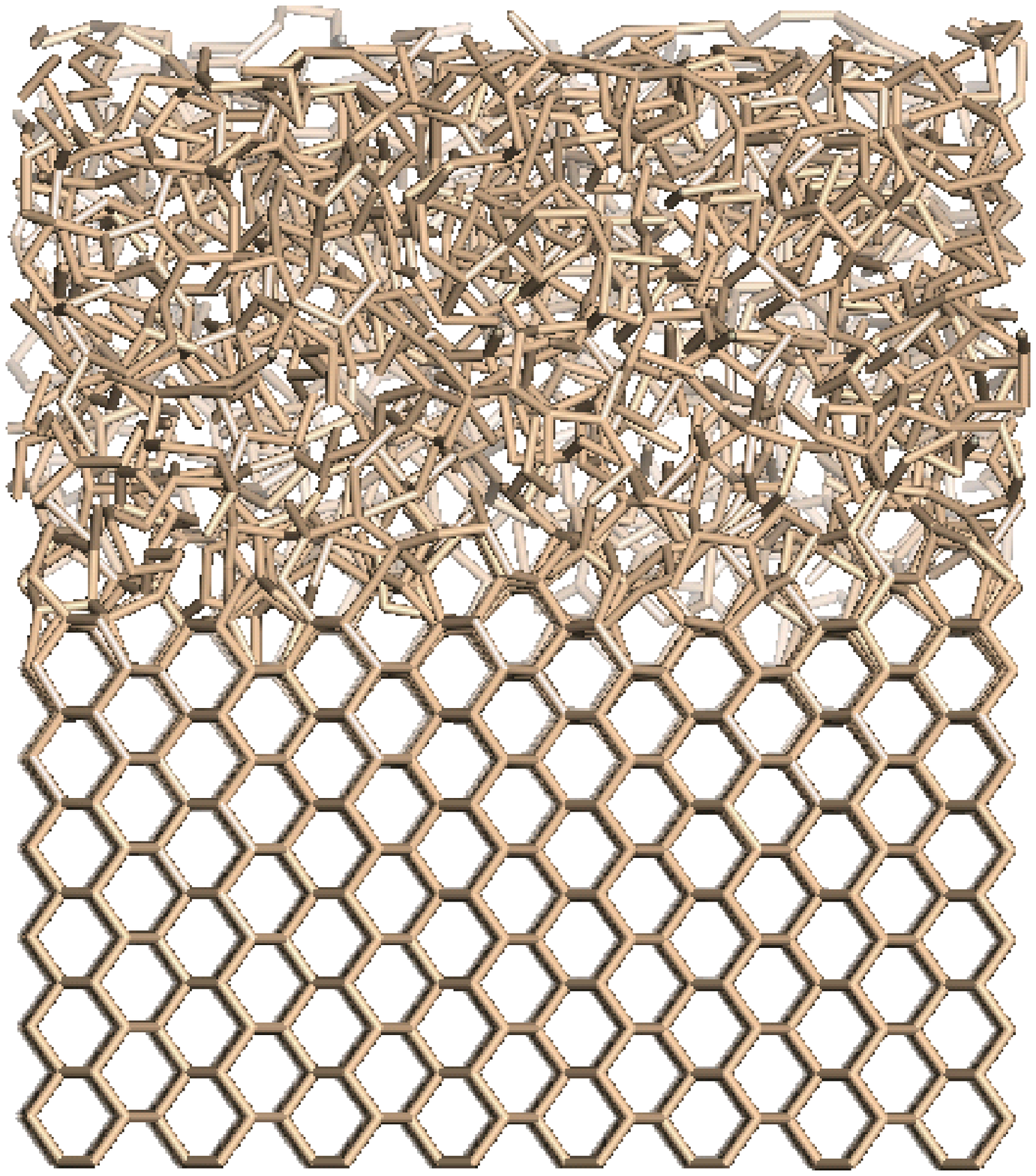}}
\resizebox{0.75\columnwidth}{!}{
\includegraphics{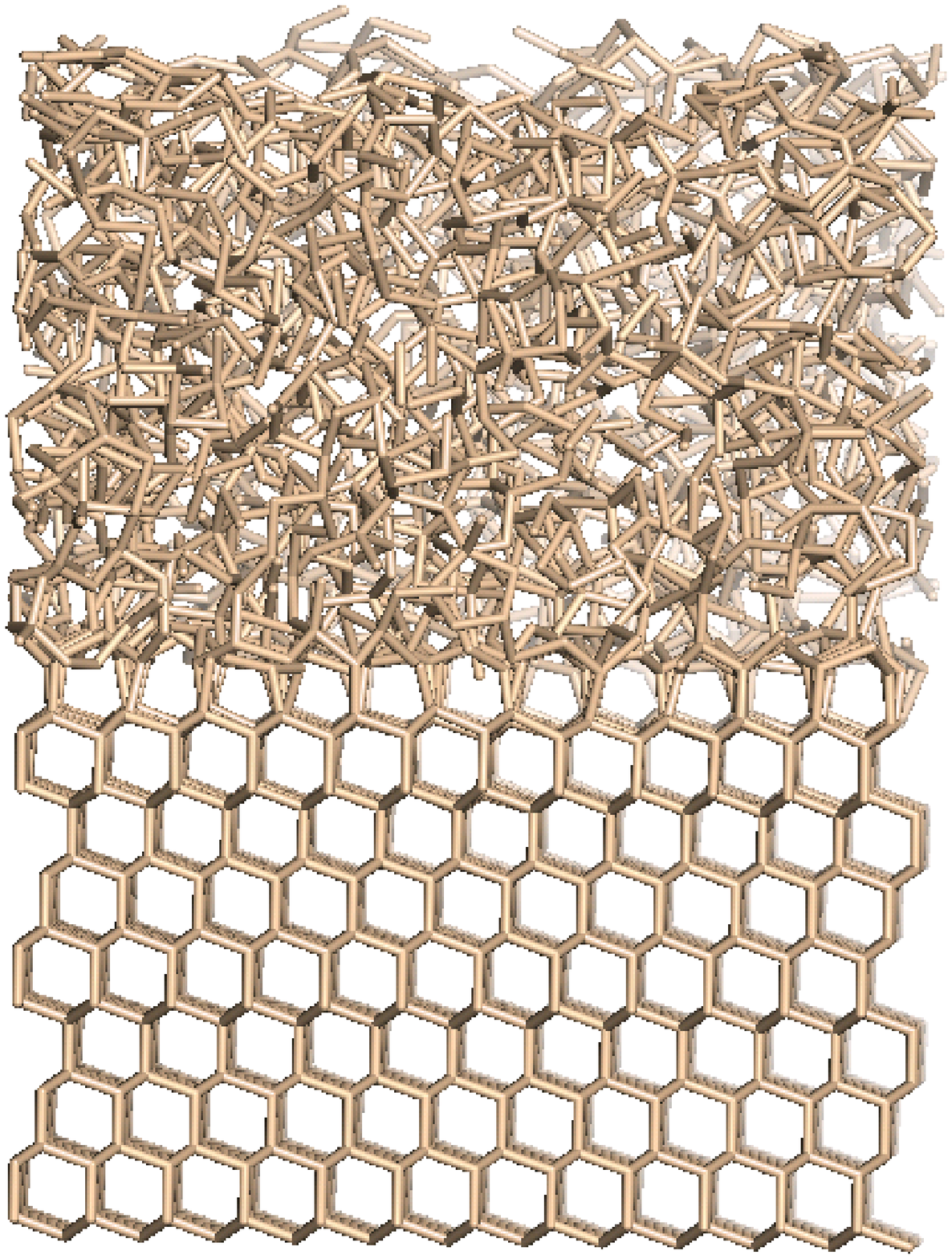}}
\caption{Stick models of the different  $a$-Si/$c$-Si interface for the three orientations $[100]$, $[110]$, $[111]$.}    
\label{fig:first_figure}
\end{figure}

\begin{figure}[tbp]
\centering

\resizebox{0.75\columnwidth}{!}{
\includegraphics{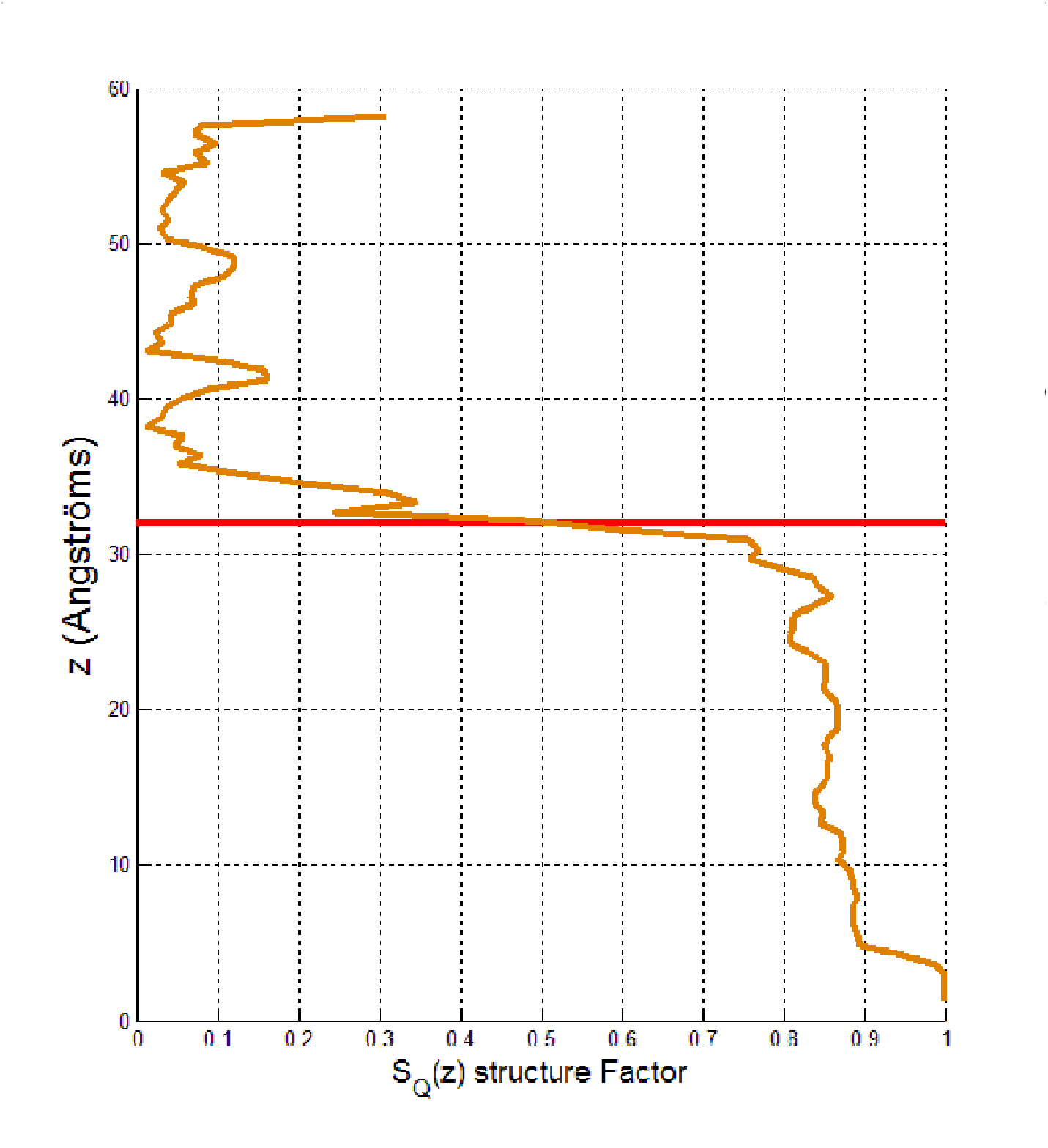}}
\caption{Example of in-plane structure factor indicating the location of the $a$-Si/$c$-Si interface. The horizontal line  at $S_{Q}=0.5$ highlights the location of the $a$-Si/$c$-Si.}
\label{fig:second_figure}
\end{figure}

Empirical molecular dynamic simulations have been undertaken in this work in order : i) to generate the amorphous and ii)  to  simulate the annealing step and the recrystallisation of an amorphous/crystalline silicon interface. In order to generate the disordered part, a melting process  simulated by a thermal annealing at  high temperature with a  low quenching rate  \cite{Ishimaru97} is considered.   The procedure has been repeated for  the three clusters  built along   directions $[100]$, $[110]$ and $[111]$ and containing respectively 4928, 5120 and 5184 atoms. The cell size parameters (8 $\times$ 8 $\times$ 10) is a trade-off between  reasonable computational costs and a possible influence of the periodic conditions on regrowth rate.    Fig \ref{fig:first_figure} illustrates the  various initial configurations. In order to simulate the layer  re-growth, the  Molecular Dynamics (MD) technique can also  be applied to  estimate the various atomic displacements at a given temperature in the framework of a Nos\'e-Hoover  thermostat \cite{Nose-Hoover}. The two most used empirical potentials Tersoff \cite{Tersoff} and Stillinger-Weber (SW) \cite{SW} are  under study in this work. The regrowth velocity determination is based on the evolution of the $c$-Si/$a$-Si or $l$-Si interface. This location  is estimated by using a plane structure factor as a function of the depth  \cite{Mattoni04} as shown in Fig. \ref{fig:second_figure}. The  temperature range for annealing is a function of the empirical potential and is given in Table \ref{table:first_table}. Only lateral periodic boundary conditions are considered  leaving the possibility to the stack to expand or contract. In order to perform a uniform doping in the amorphous part, a specific approach has been developed. The amorphous part is sliced in various trenches of different thicknesses. Boron atoms are introduced in the amorphous lattice in a substitutional way. After the dopants introduction, the equilibration of the cluster is performed by a short annealing step leading to a final dopants distribution located in the amorphous part. Different concentration levels of boron are considered from the lowest one (0.2\%) up to (5.4-6\%). In the case of Molecular Dynamics with doped boron  systems, only  the  Tersoff potential has been used. It is often claimed   that no empirical potential describes  properly the subtle  Si-B and B-B interactions. In this work, the approach of P\'erez-Mart\'in  {\it et al.} \cite{Perez_Martin04} is followed where boron atoms are assumed to adopt a similar behaviour in silicon like carbon atoms.

\subsection{Recrystallisation velocity and disordered part  properties}

\begin{figure}[tbp]
\centering

\resizebox{1\columnwidth}{!}{
\includegraphics{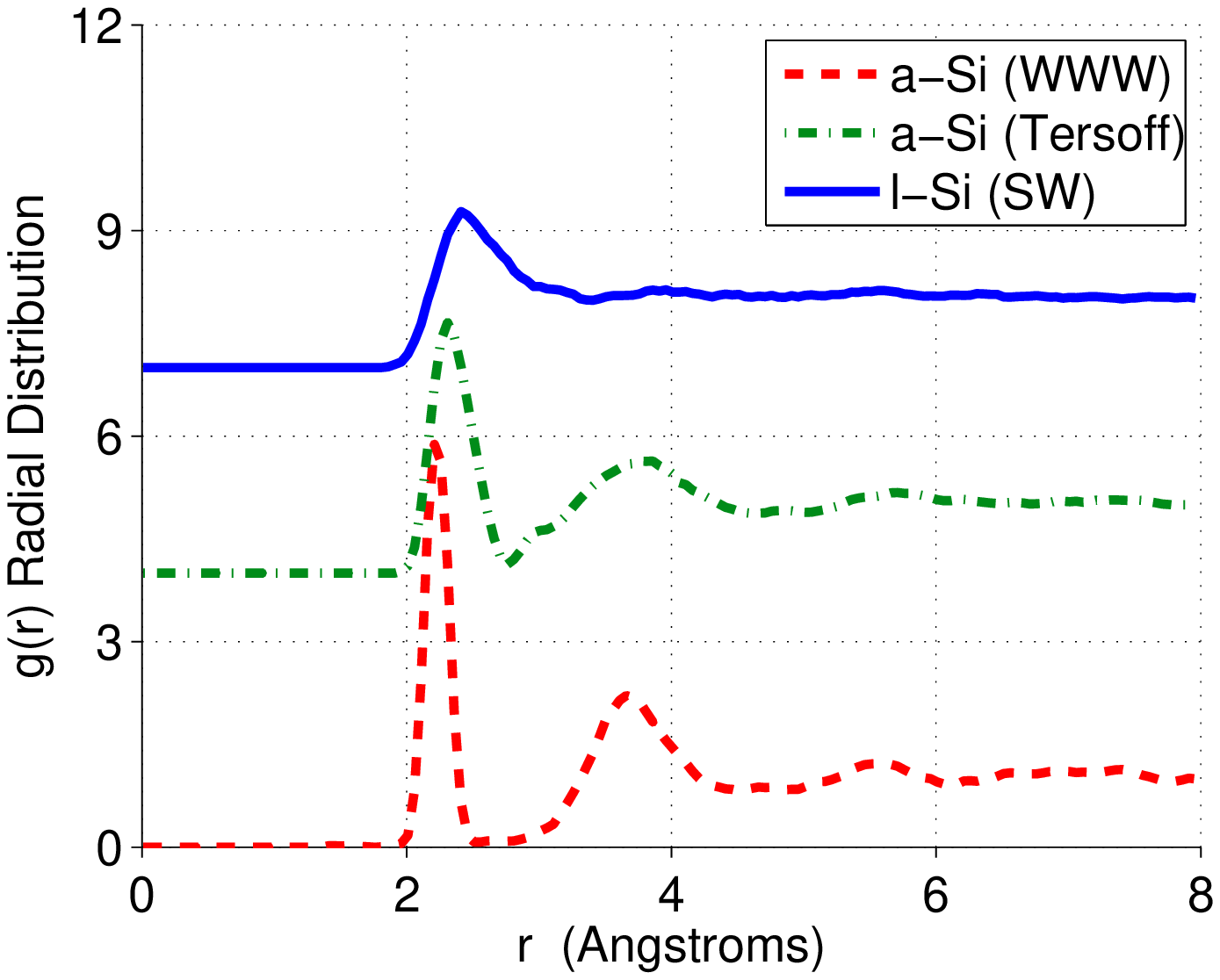}}
\resizebox{1\columnwidth}{!}{
\includegraphics{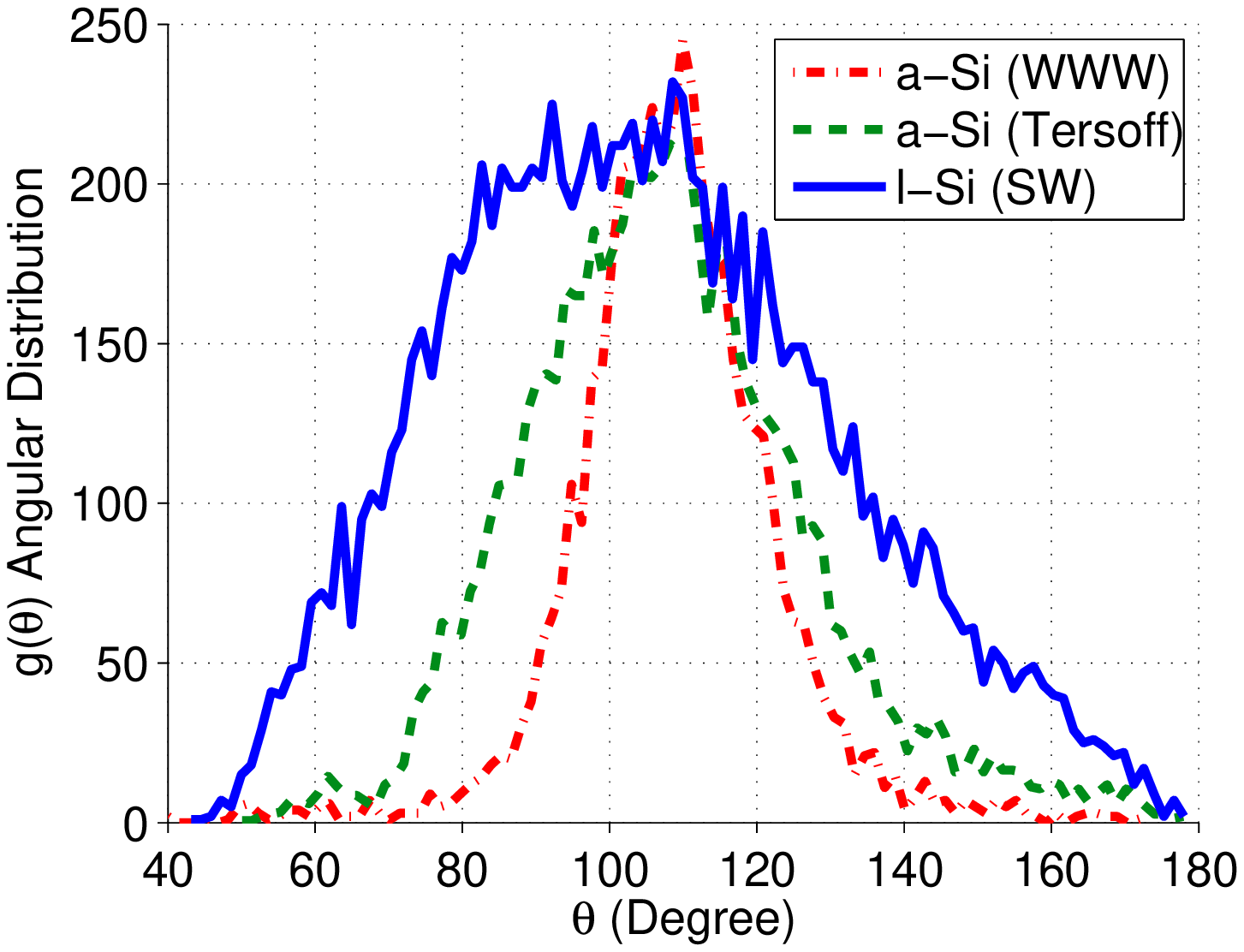}}
\caption{Radial and angular distribution properties for the different potentials and the amorphous reference generated by the Wooten Weiner Weaire approach.}
\label{fig:third_figure}
\end{figure}

\begin{figure}[btp]
\centering

\resizebox{1\columnwidth}{!}{
\includegraphics{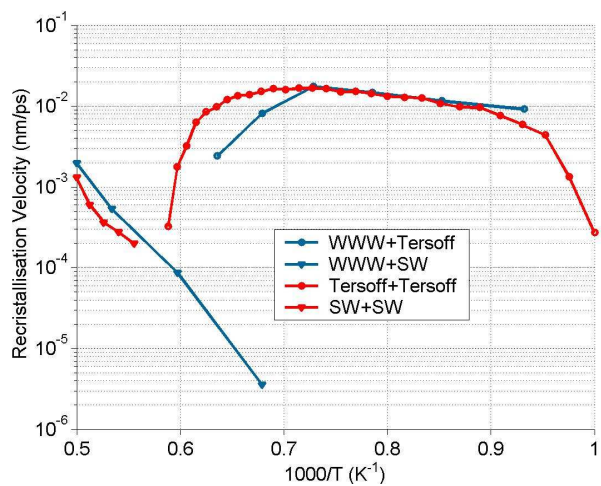}}
\caption{Evolution of the $[100]$ interface re-growth  velocity as a function of the annealing temperature  for the two main types : Wooten Weiner Weaire (WWW) approach and melting/quench and for the two main potentials Tersoff and Stillinger-Weber (SW).  The variation remains relatively modest compared to the  difference in atomic cluster size (2000 atoms for WWW) and this work ($\sim$ 5000 atoms) and those with  the initial interface structure.}
\label{fig:four_figure}
\end{figure}

In a previous work, it has been shown that for the  simulation  of silicon solid phase epitaxy  the empirical potential must be  taken  carefully into account with  the high sensitivity to describe or preserve an amorphous structure during annealing \cite{Krzeminski07}. However, there is a  debate  on  the influence of the initial properties of the disordered part and  the recrystallisation velocity obtained by molecular dynamics. In this work,  the quality of the initial interface has been enhanced in comparison to previous studies  where  only a Wooten Winer Weaire's amorphous cluster on a crystalline box was considered. The two main usual methods for generating a virtual  amorphous cluster (Monte-Carlo bond switching \cite{Wooten85} or by melting-quench) are confronted. The melting temperature used to generate a liquid part  and next  the quenching rate is described in Table \ref{table:first_table}.  As shown in Fig. \ref{fig:third_figure} and also in table \ref{table:second_table}, large differences for the  different properties (from radial and angular distribution to  the mean coordination number distribution) in the disordered part can be observed. Fig. \ref{fig:four_figure} reports the recrystallisation velocity of the $a$-Si or $l$-Si/$c$-Si interface for the two main potentials Tersoff and Stillinger-Weber (SW). For the recrystallisation velocity obtained   using a  Tersoff potential,  Arrhenius kinetics have been observed for the  two different  amorphous in relative close agreement  characterised with the same  activation energy of $\sim$ 2.9 eV.  The situation is completely different in the case of a simulated annealing using Stillinger-Weber. Despite the fact that in the first case, an initial $a$-Si/$c$-Si  structure whereas in the second case a completely different $l$-Si/$c$-Si  interface  is under consideration, small differences can be observed.  Clear liquid phase epitaxy characteristics with velocity re-growth of about a few $\sim$  m/s in a wide  temperature range limited by the melting of the crystalline part  is observed slightly above a temperature of 1675K. These results strengthen our previous conclusions about the major influence of the empirical potential used for the estimation of the recrystallisation velocity and on the variability of empirical potentials to preserve an amorphous structure upon annealing \cite{Krzeminski07}.

\section{Results and discussion}

\subsection{Orientation effects and solid phase epitaxy}
\subsubsection{Regrowth mechanisms}

\begin{figure}[btp]
\centering

\resizebox{1\columnwidth}{!}{
\includegraphics{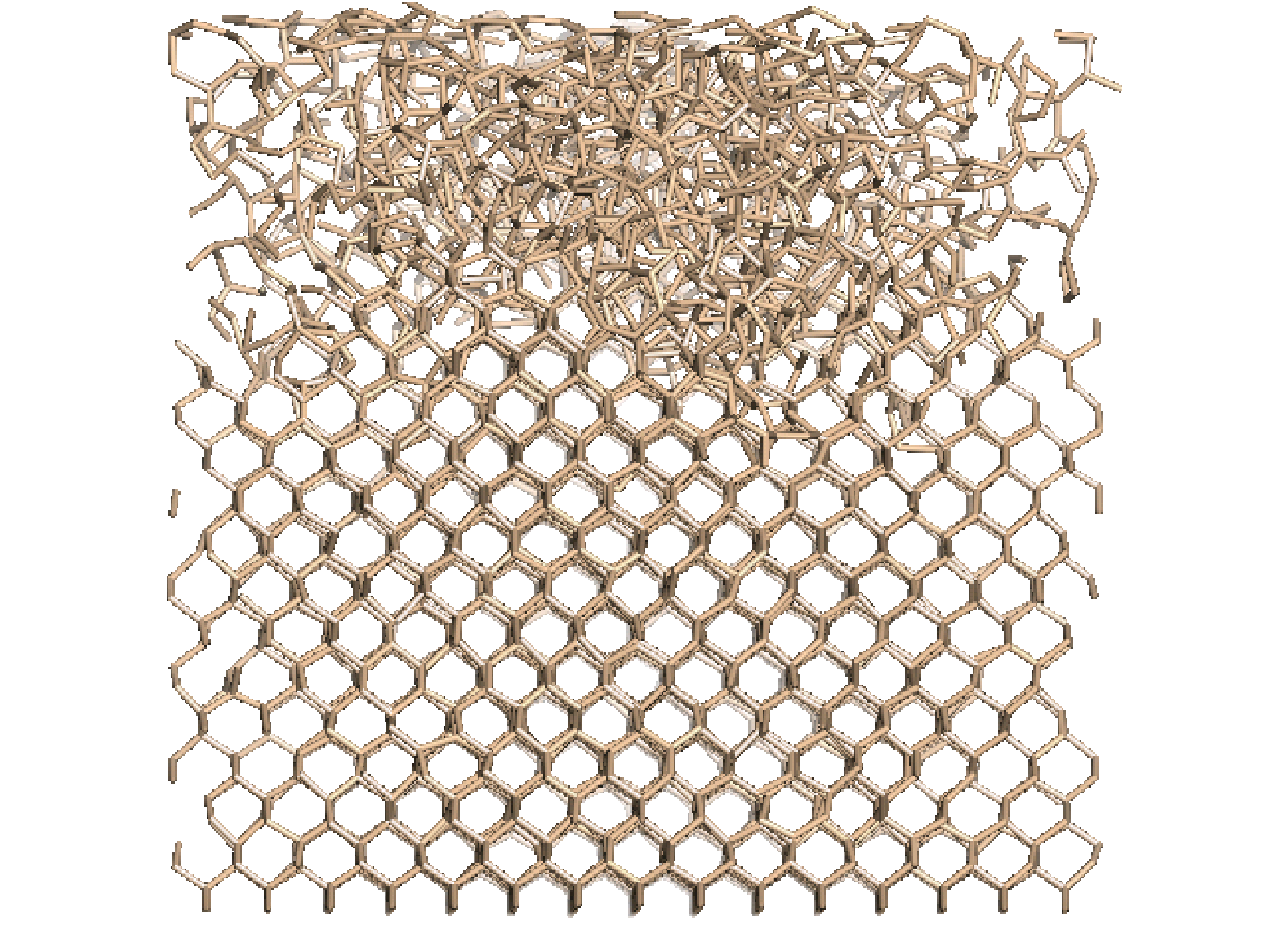}
\includegraphics{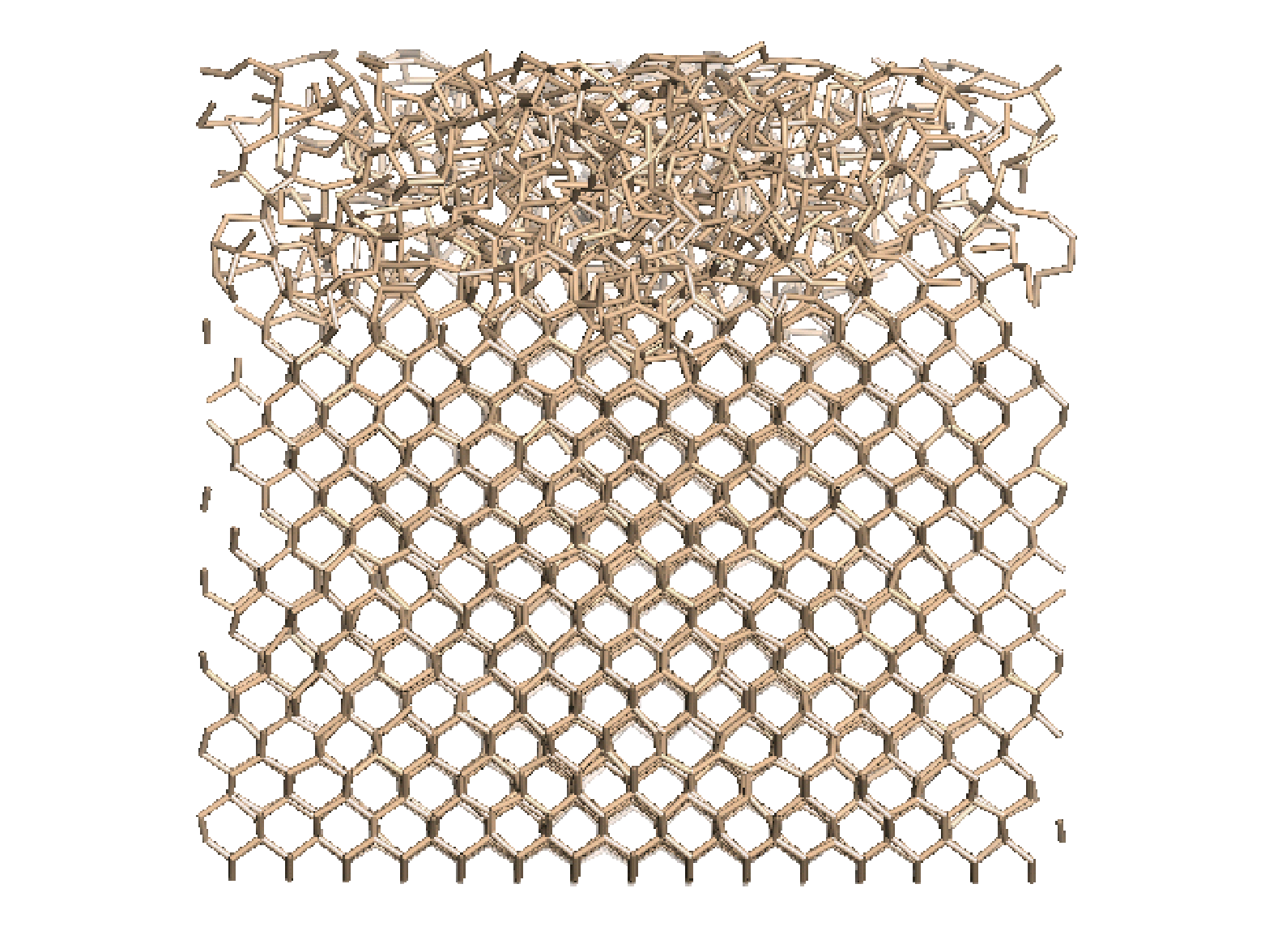}}
\resizebox{1\columnwidth}{!}{
\includegraphics{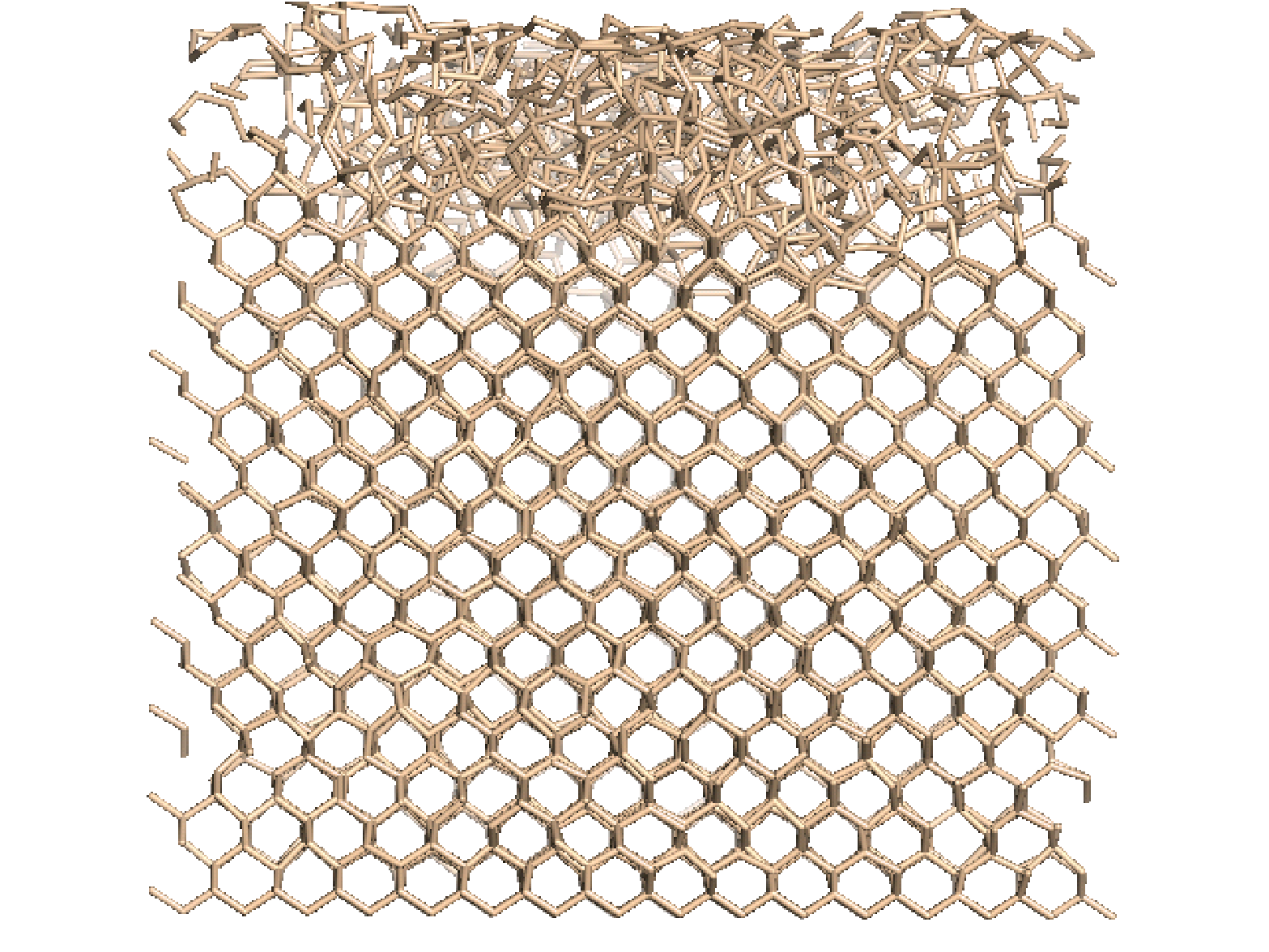}
\includegraphics{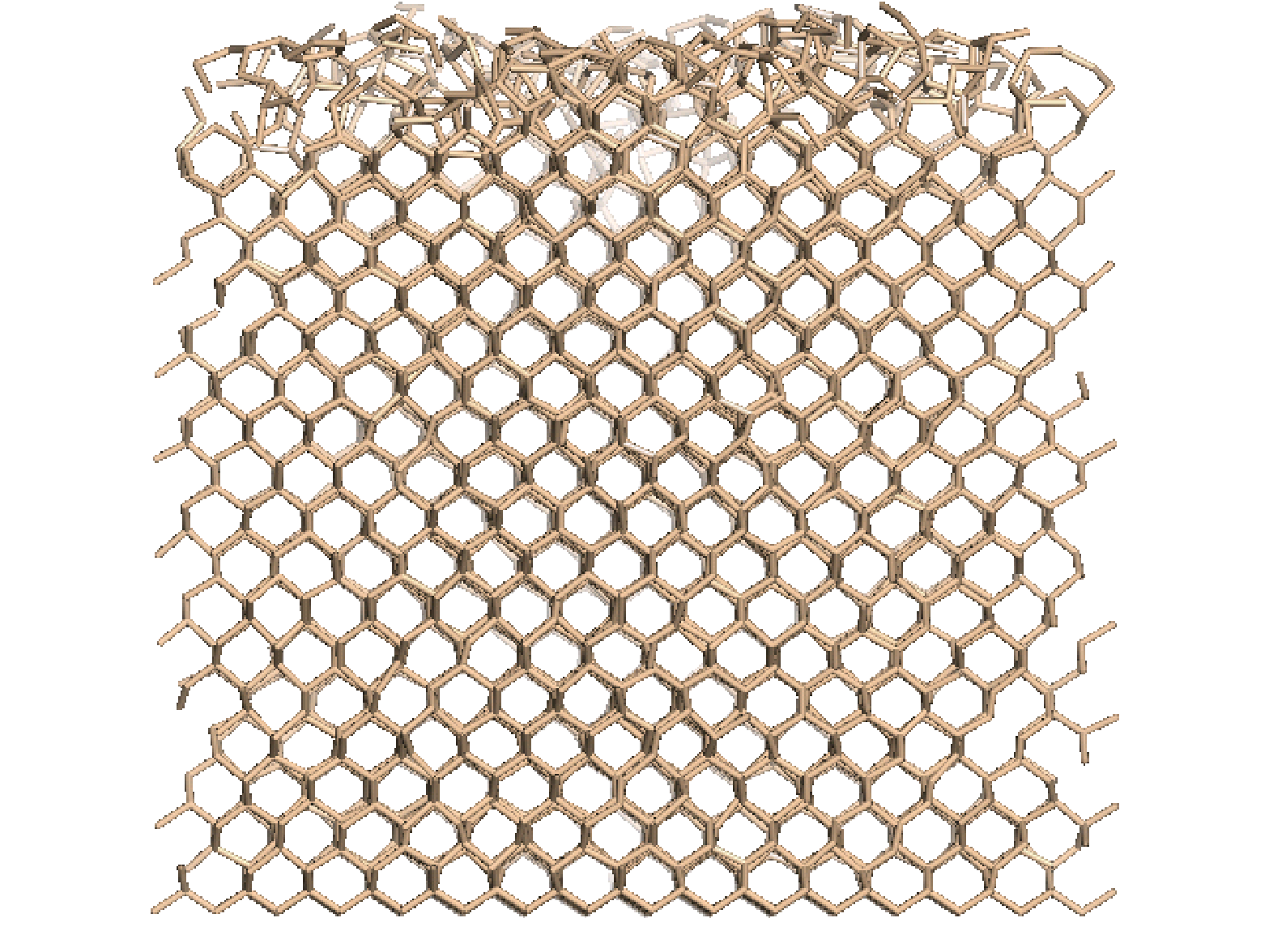}}
\resizebox{0.5\columnwidth}{!}{
\includegraphics{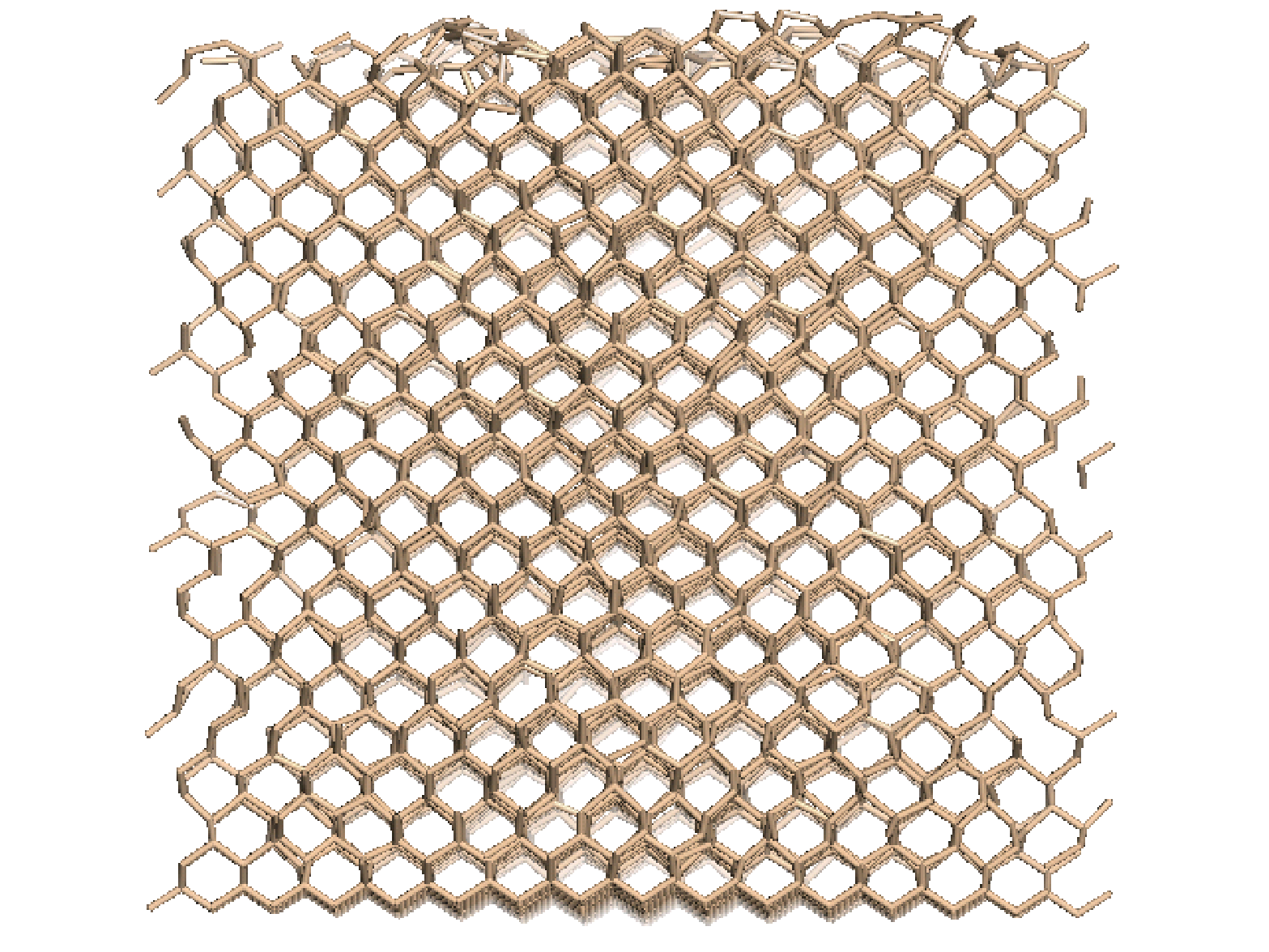}}
\caption{Evolution of the $a$-Si/$c$-Si $[100]$ interface position  during solid phase epitaxy at 1700K. The evolution is reported for respectively 2.5, 5, 10, 15 and 20 ns illustrating the recrystallisation process where a quasi-planar interface propagates by a one atom by one atom incorporation mechanism.}
\label{fig:five_figure}
\end{figure}

The influence of the interface orientation effect on the re-growth mechanism has been investigated.  The interface migration process is studied for the different crystalline orientations  $[100]$, $[110]$ and $[111]$. Fig. \ref{fig:five_figure}  presents the $a$-Si/$c$-Si  for an interface migration at 1700K during thermal annealing. For $[100]$, the re-growth front is governed by first a diffusion process  in the  disordered part, then by a partial ordering at the interface and finally by a one-by-one  atom incorporation. The propagation of the crystalline front proceeds in  a quasi-planar  $a$-Si/$c$-Si interface with some partial reordering which could extend to one or two planes. 

\begin{figure}[btp]
\centering

\resizebox{1\columnwidth}{!}{
\includegraphics{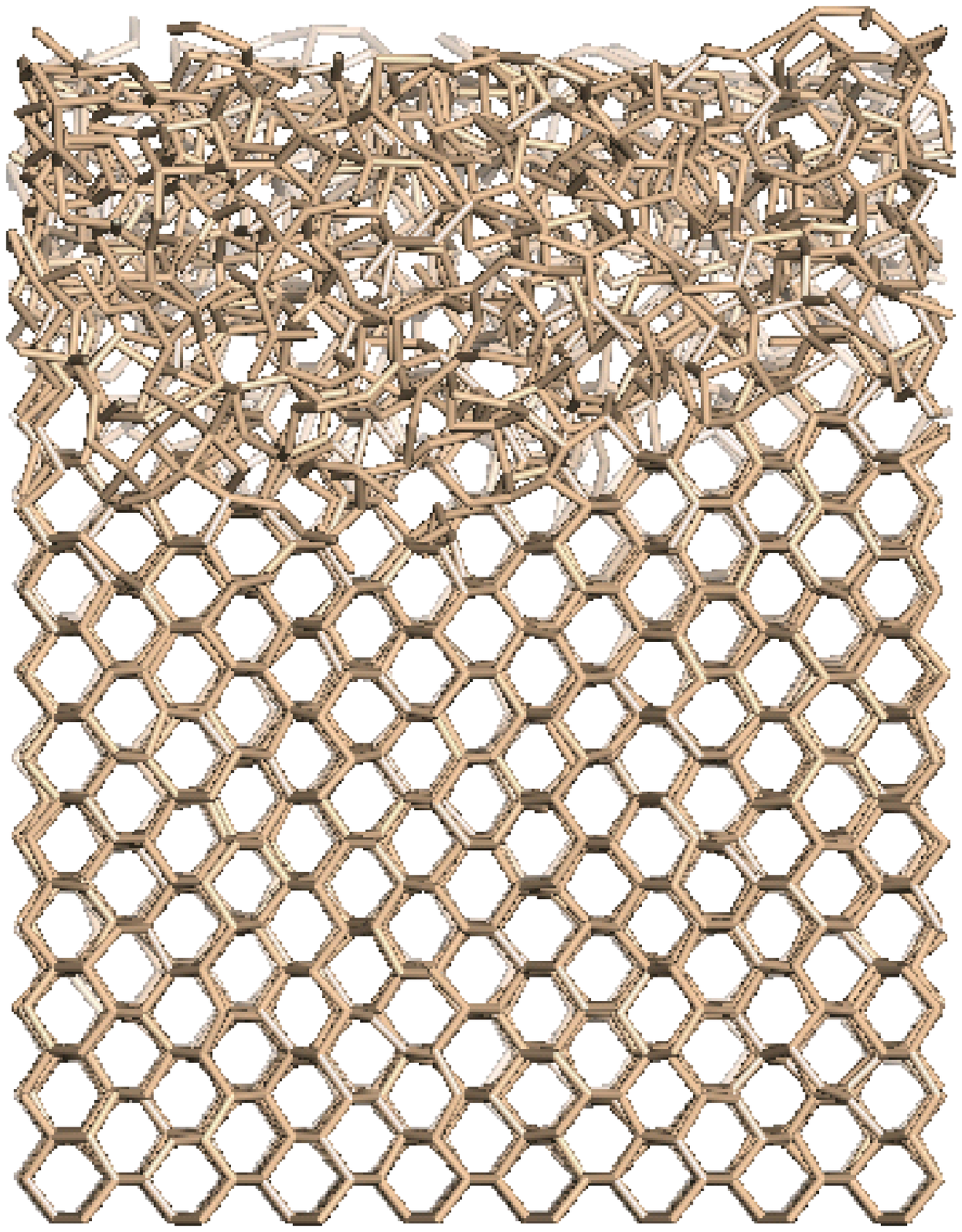}
\includegraphics{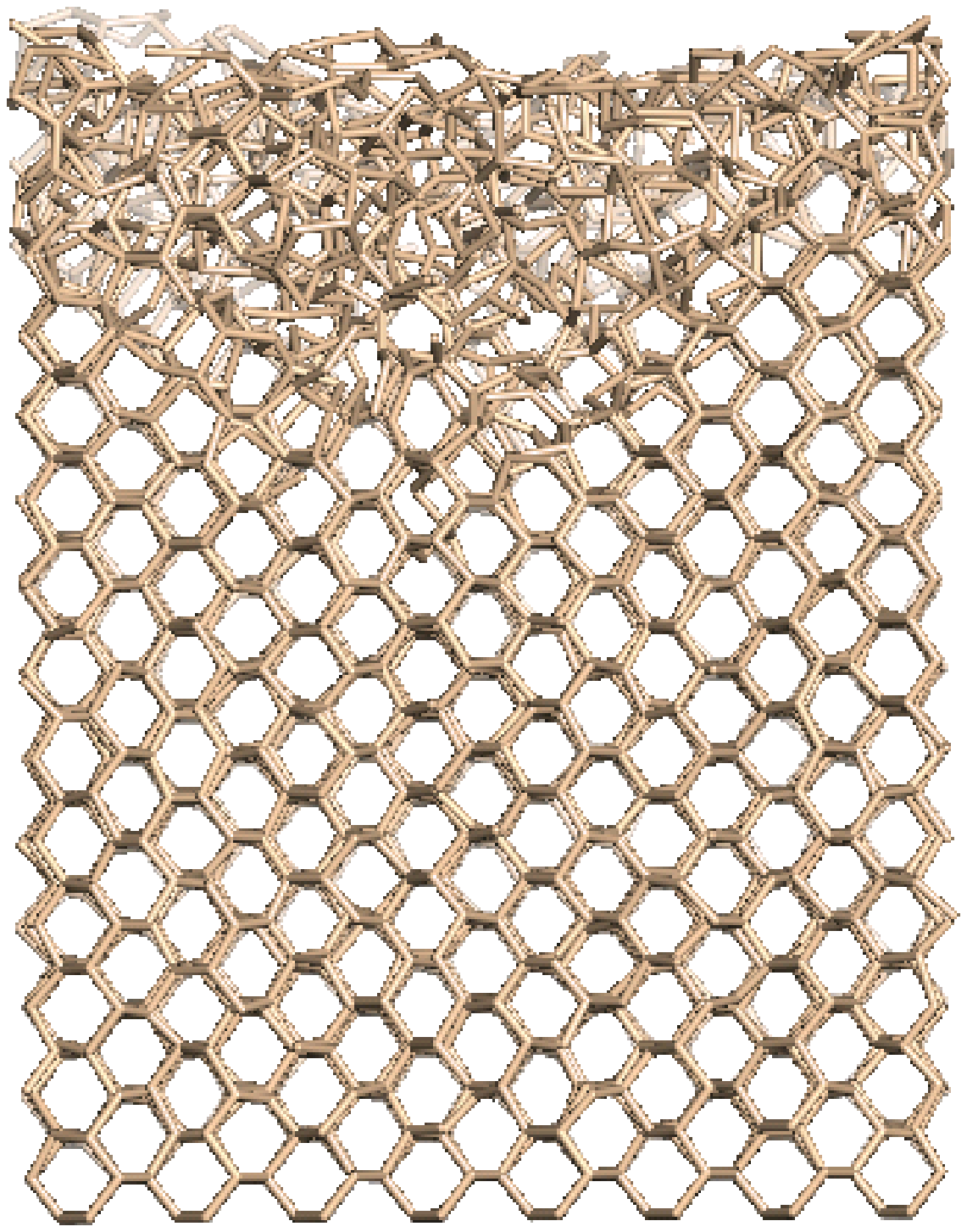}
}
\resizebox{1\columnwidth}{!}{
\includegraphics{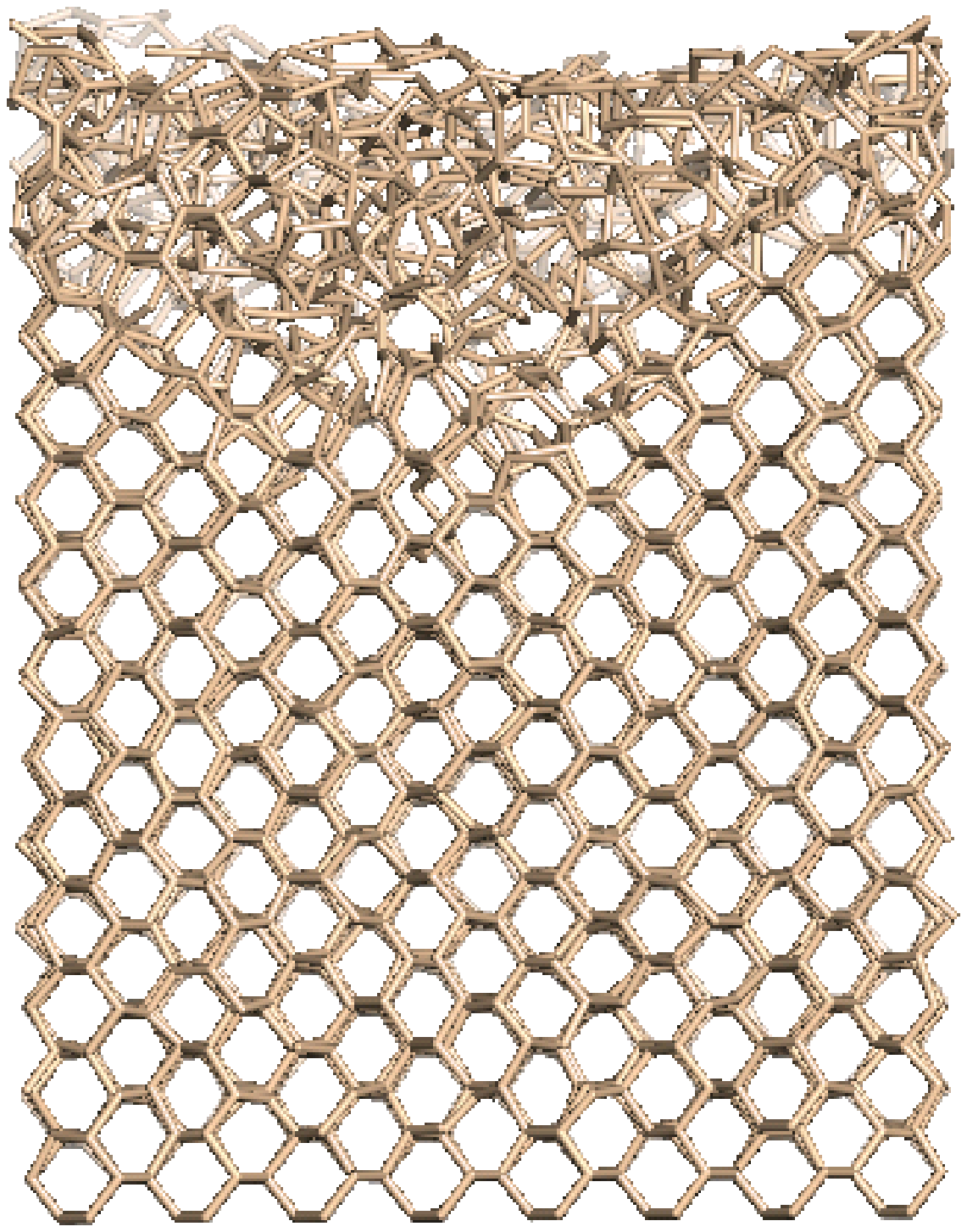}
\includegraphics{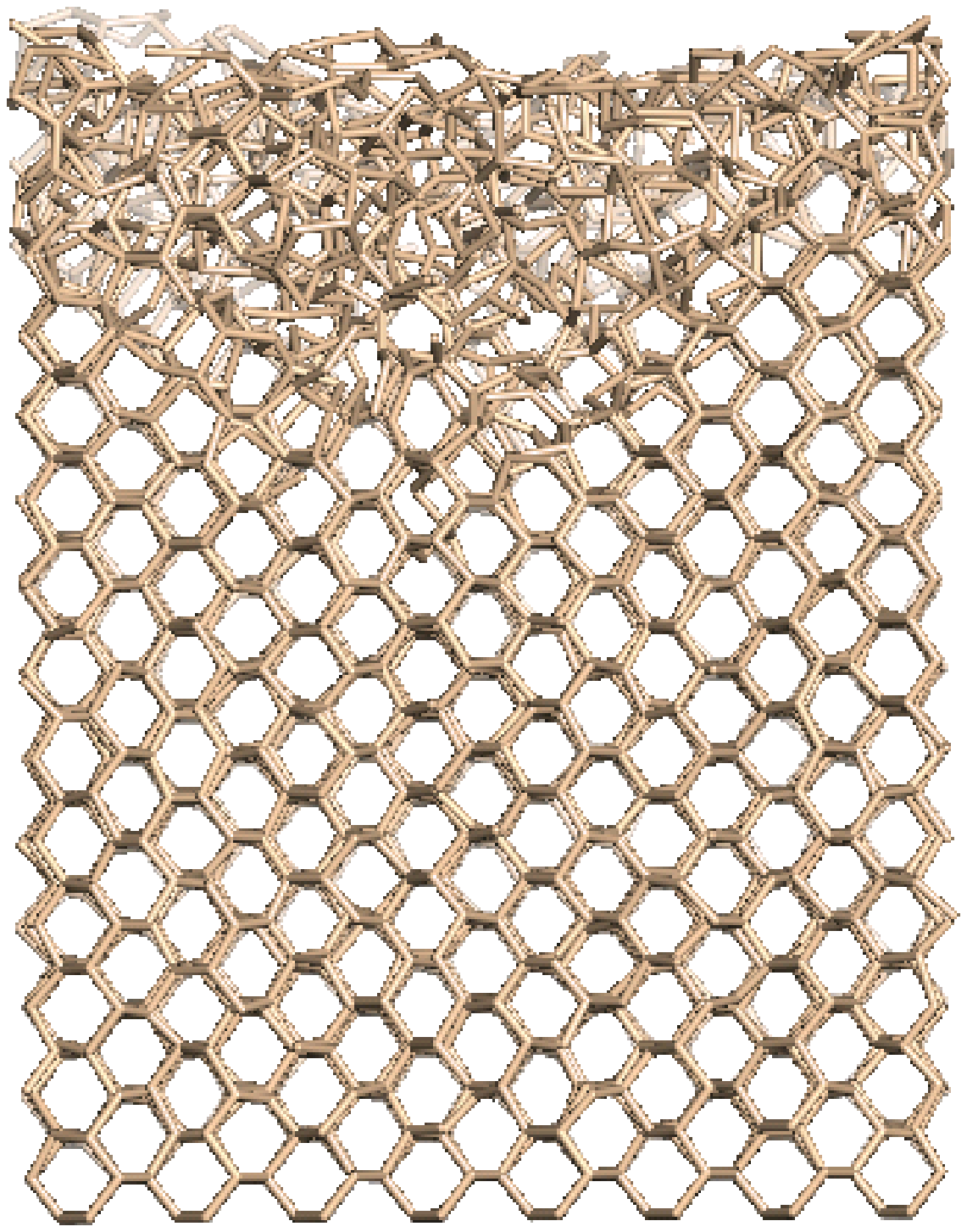}
}
\resizebox{0.5\columnwidth}{!}{
\includegraphics{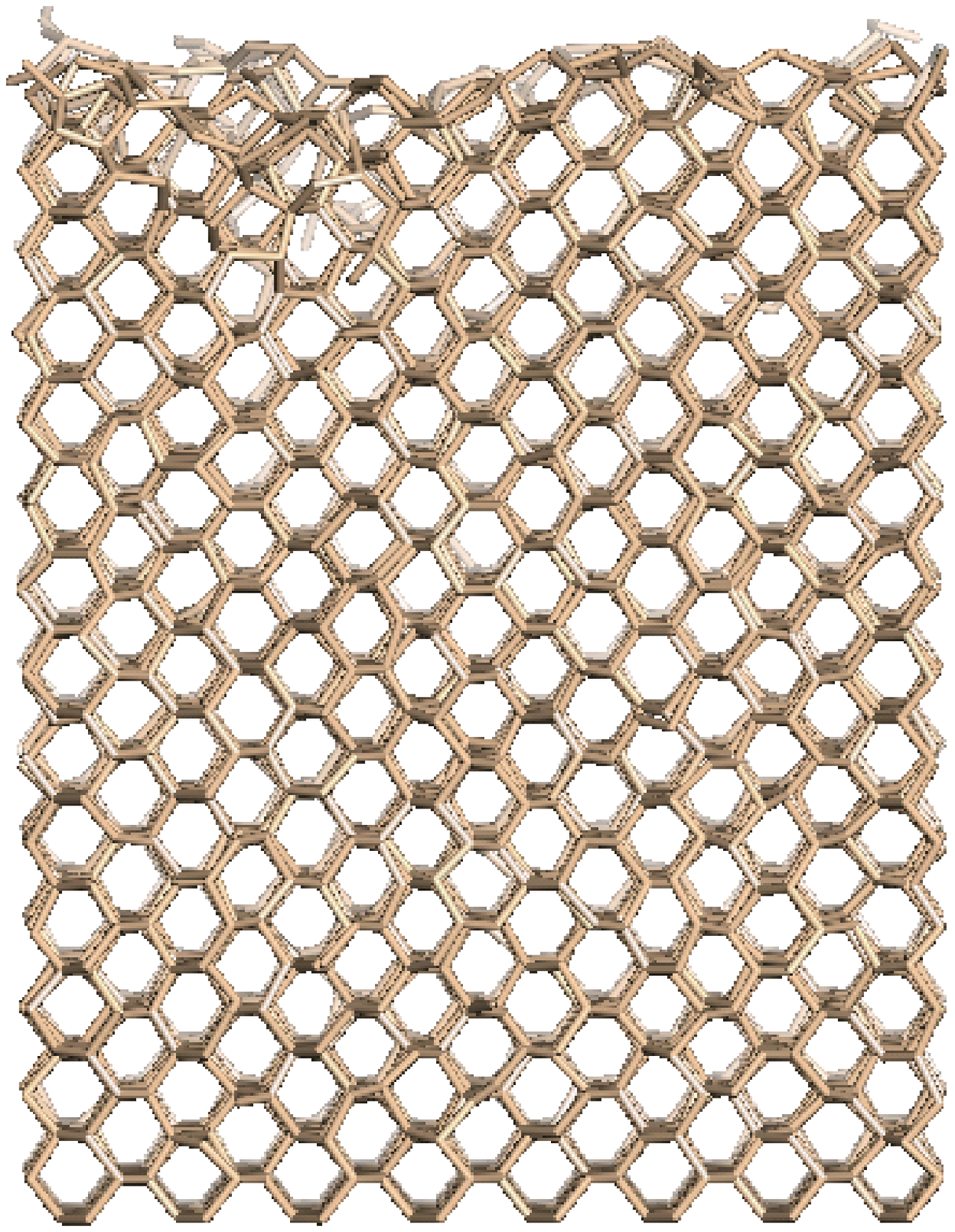}
}
\caption{Evolution of the $a$-Si/$c$-Si $[110]$ interface position  during solid phase epitaxy at 1700K. The evolution is reported for respectively 5, 10, 15, 20  and 25 ns  illustrating the  recrystallisation mechanism where the formation  and the dissolution of nanofacets  is observed.}
\label{fig:six_figure}
\end{figure}

For the orientation  $[110]$ a relative different mechanism  has been observed in Fig. \ref{fig:six_figure}.  Two silicon atoms have to be incorporated at the recrystallisation front  and then to bind with  each other. In order to get  a second atom and to complete the crystalline plane, the recrystallisation is observed to propagate also vertically in another nucleation site. The re-growth front predicted by simulation is made of nanofacets leading to the formation of a sawtooth interface of about 3-4 atomic planes. For $[110]$,  the interface is more diffuse and characterised by a more important roughness as a partial ordering extend to several atomic planes. 

\begin{figure}[btp]
\centering

\resizebox{0.7\columnwidth}{!}{
\includegraphics{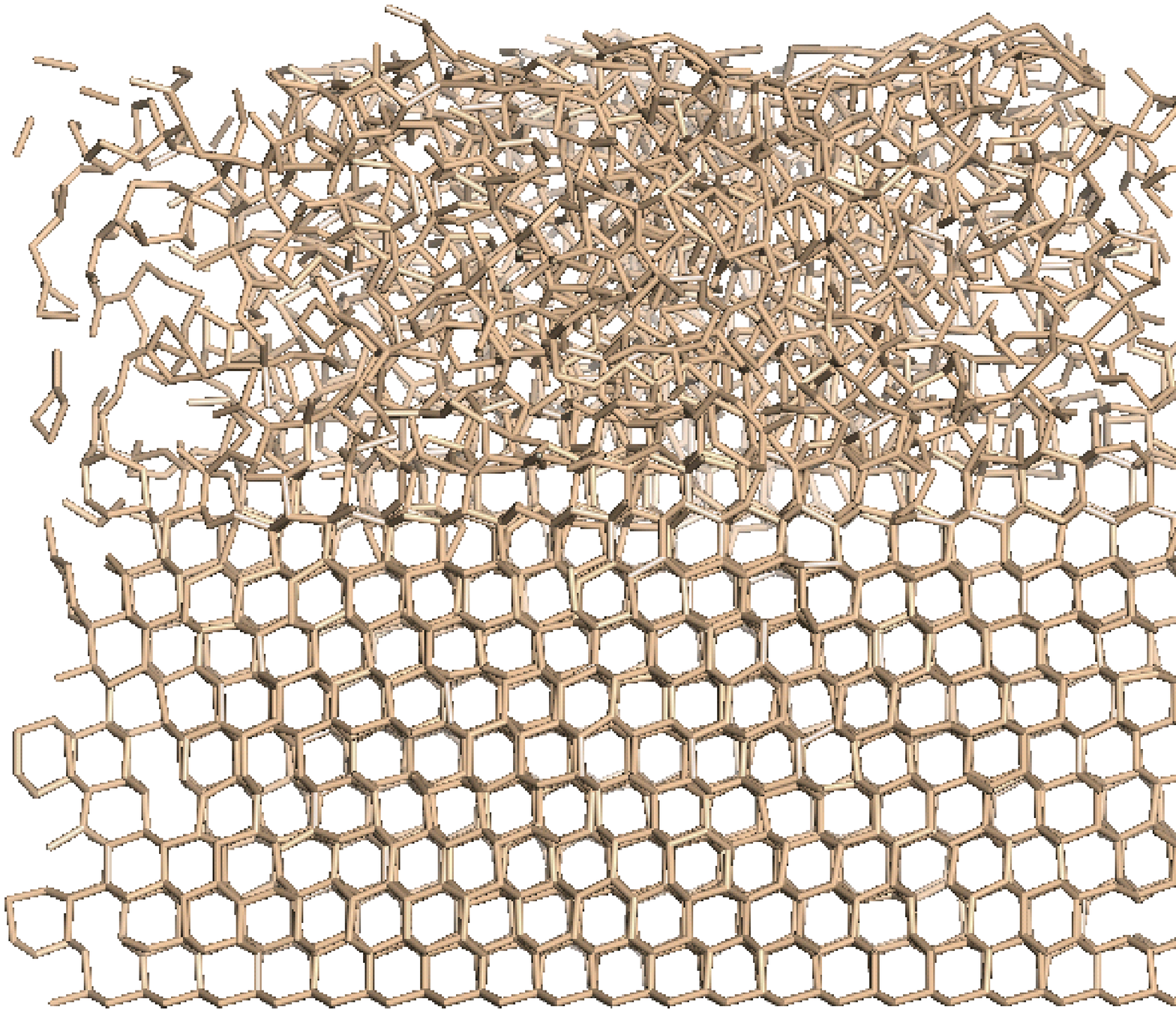}}
\resizebox{0.7\columnwidth}{!}{
\includegraphics{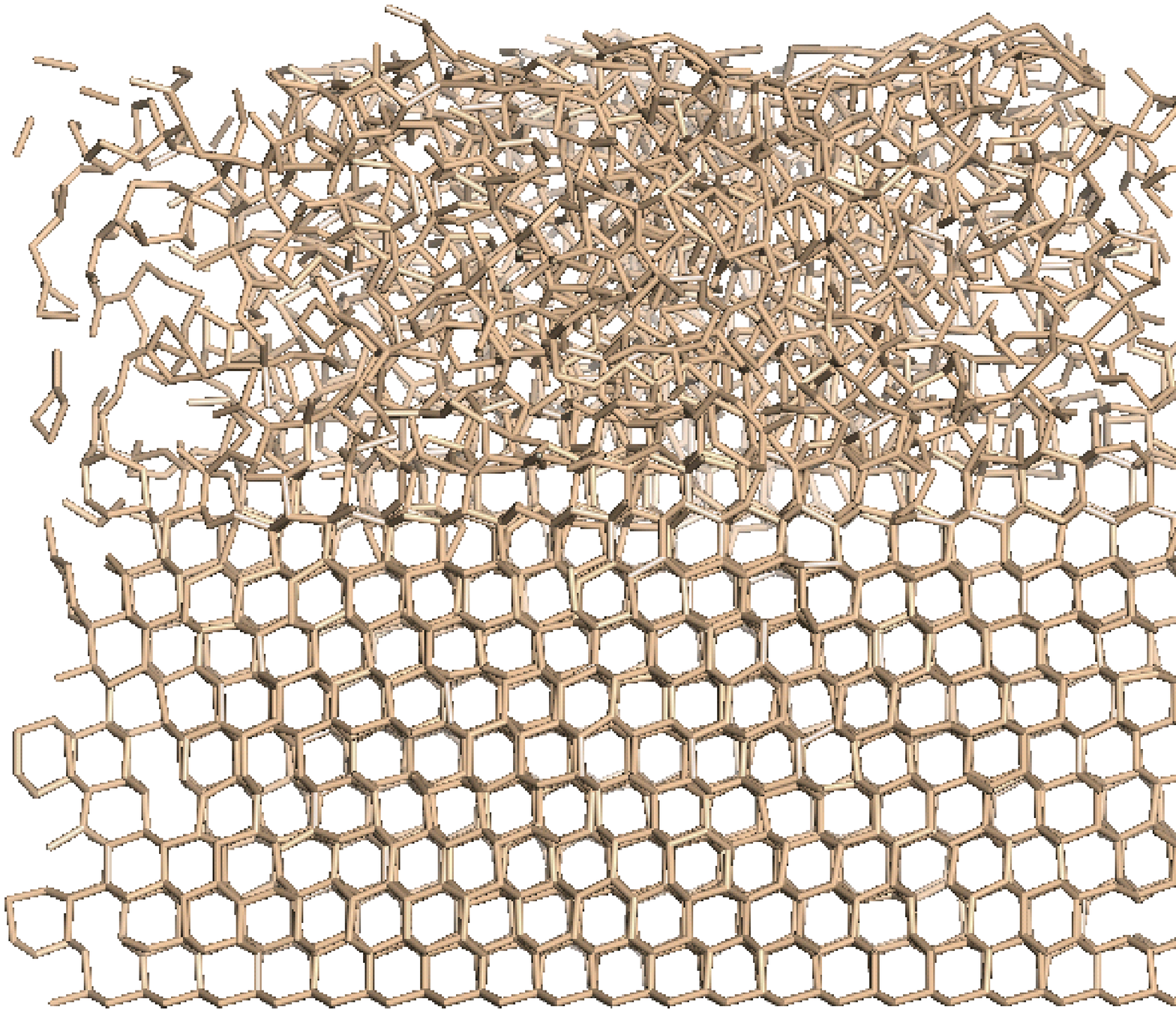}}
\resizebox{0.7\columnwidth}{!}{
\includegraphics{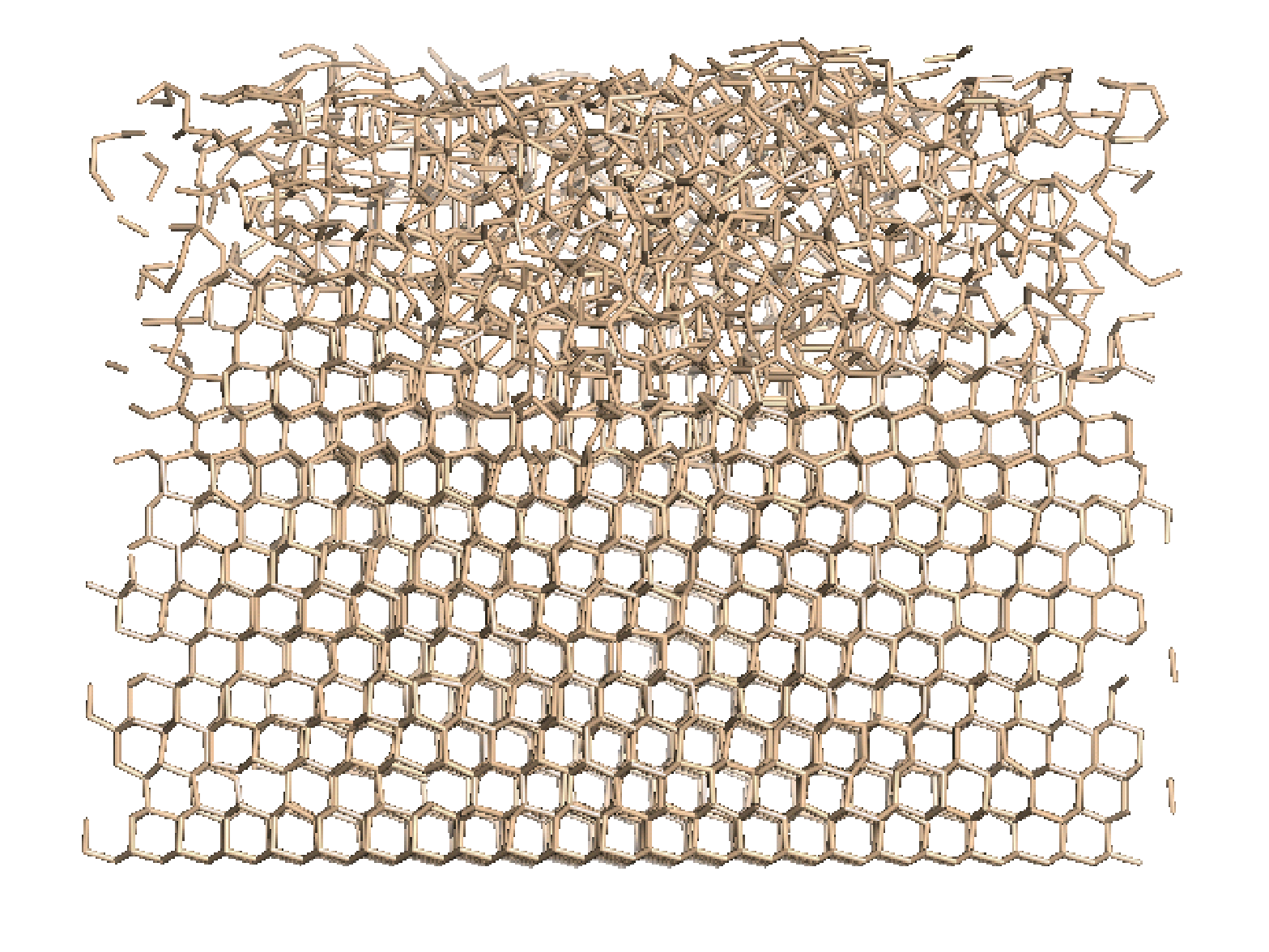}}
\resizebox{0.7\columnwidth}{!}{
\includegraphics{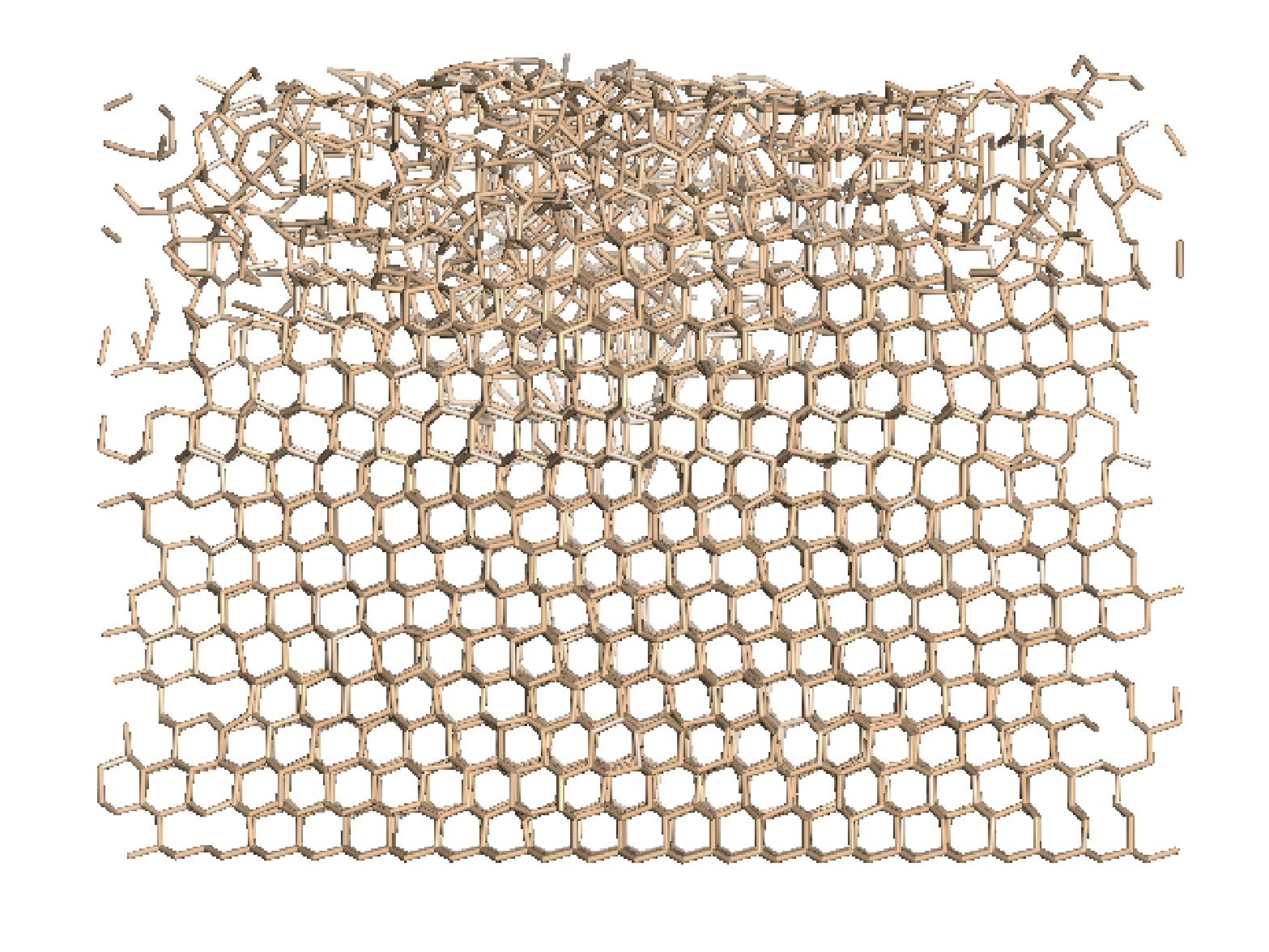}}
\caption{Evolution of the $a$-Si/$c$-Si $[111]$ interface position  during solid phase epitaxy at 1700K. The evolution is reported for respectively 5, 10, 15, and 20 ns  illustrating the bilayer  recrystallisation mechanism and defect formation during re-growth.}          
\label{fig:seven_figure}
\end{figure}

For the  case  $[111]$, the recrystallisation mechanisms are observed to be even more complex (Fig. \ref{fig:seven_figure}) in comparison with $[100]$ and $[110]$. The migration from $a$-Si to $c$-Si occurs by the rearrangements of complete or partial bilayers. For example, in the case of Fig. \ref{fig:seven_figure} incomplete bilayers  form on the right  and left  part, while the central part is slowly rearranging (10 ns and 20 ns) letting suddenly appear several crystalline layers (25 ns). A buried defect at the top surface remains.  The  interface  migration is made of successive plateaus whose height correspond to the distance (or a multiple of the distance) between  the  bilayers $(111)$. The formation of one or more twins in planes $(111)$  has been observed in nearly all the temperatures simulated. Other defects are observed in  about 50~\% of the simulations. The $[111]$ interface recrystallisation interface is also completely different and  mostly based on a  bilayer process. For $[111]$, the silicon atoms tends to organise partially in bilayer with a significant time necessary for the atomic arrangements within the the bilayer as three atoms of the bilayer are involved in the atomic process for the creation of a six-fold rings.  Again, in the case  $[111]$, the recrystallisation process is not strictly planar with a large interplay between  in  and out of plane nucleation.

\subsubsection{Orientation dependent recrystallisation velocity}

\begin{figure}[btp]
\centering

\resizebox{1\columnwidth}{!}
{\includegraphics{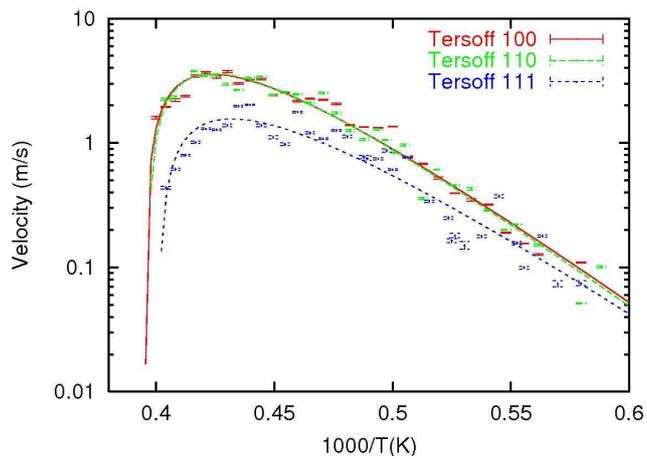}}
\caption{Evolution of the interface recrystallisation velocity along orientation $[100]$, $[110]$ and $[111]$  for the Tersoff potential. The orientation influence concerns mostly the prefactor since the activation energy of the regrowth process is observed not to be  orientation dependent.}  
\label{fig:eight_figure}
\end{figure}

One main open question concerns the influence of the re-growth mechanisms related to orientation effects  at the  $a$-Si/$c$-Si interface.  Fig. \ref{fig:eight_figure} presents an example of  the interface velocity extraction in the intrinsic  case for the three orientations under study. This corresponds to the case where it was possible to  define an $a$-Si/$c$-Si interface using a planar structure factor. This is not possible for in-plane and extended defects which generate too much  disorder to allow a proper interface definition. All the systems considered  are free from defects   except for  systems $[111]$ containing only twin parallel defects. In this particular case,  the orientation $[111]$  is estimated to be kept since it is only  a default in the stacking plane sequence. The results as a function of the temperature and for the three orientations are given in Fig. \ref{fig:eight_figure} in the temperature range that could be considered with a reasonable computing time. It can be observed that  while being very close in the orientations  $[100]$ and $[110]$, the velocity is lower along the  direction $[111]$ by a factor 1.43. For the  orientation $[100]$, the activation energy  can be extracted using the classical theory of thermally activated growth \cite{Lu91}.  Considering the fact that the interface velocity  reported in the Arrhenius plot is superimposed or either parallel,  there was no reason to consider a different activation energy  for the  cases  $[110]$ and $[111]$.

\subsection{Discussion}

The theoretical picture given by molecular dynamics simulations stresses the importance of the initial interface orientation on the growth mechanisms. It has been observed that solid phase epitaxy mechanisms are strongly dependent on  the  interface orientation with a quasi-planar migration for the $[100]$ interface. This  propagation is compatible with most analytical modelling based on a  sharp interface migration. For $[110]$ and $[111]$, the situation is relatively different with the formation respectively of  nanofacets for $[110]$  and a bilayer mechanism leading to the formation and the migration of a  more complex and  diffuse interface  associated with the generation of  defects. Even if the physical reality  described by molecular dynamics is more complex, some connections could be made with the phenomenological model of Drosd and Washbrun \cite{Drosd82}  which described the initial recrystallisation steps \cite{Lampin09}. The results on the interface recrystallisation velocity are more complicated to confront with experimental results. One main issue in our study is to define an interface migration velocity in the case where the $a$-Si/$c$-Si interface is more diffuse and  characterised by  the presence of  defects. In this paper, only free-defect re-growth is considered. It  corresponds to an ideal physical picture which is not observed experimentally \cite{Drosd82}.  The resulting activation energy of the process for the various orientation is 2.87 eV in accordance with the experimental range of [2.4-2.9]eV \cite{Drosd82,Olson88,Csepregi78}. This activation energy  cannot  be associated with a simple active process like self-diffusion or defects migration. This energy seems rather  associated with the energy barrier at the interface as suggested by Posselt {\it et al.} \cite{Posselt09}.  Since some bond breaking and exchange between the amorphous and the silicon phase and exchange are often observed in our simulation in order to allow a proper definition and propagation of the $a$-Si/$c$-Si interface. Some links could be made with the picture given by molecular dynamics and the approach of Lu {\it et al.} \cite{Lu91} which  considered the dangling bond breaking, moving and trapping mechanism as the main process responsible for the growth rate limitation \cite{Aziz91}.  The prefactor for orientation  $[111]$ (1.2$\times$10$^{7}$ m/s)  is reduced by a factor 1.43 compared to $[100]$ and $[110]$ suggesting that the  atomic incorporation mechanism based on three atoms at interface [111]  slows down the overall  interface migration  compared to the other orientation. Only a small part of the velocity reduction can be explained by our molecular dynamics simulation. Since experimentally more interface recrystallisation velocity reduction is observed for $[110]$ and $[111]$, the role  and the influence of defects on the nucleation velocity should be investigated more in depth at the theoretical level. A remaining open question that should be addressed concerns the influence of the cell size particularly in a large  presence of defects. It seems to be the most plausible hypothesis to explain the reduction velocity effects.

\subsection{Regrowth of a doped $a$-Si:B/$c$-Si interface}

\begin{figure}[btp]
\centering
\resizebox{0.4\columnwidth}{!}{\includegraphics{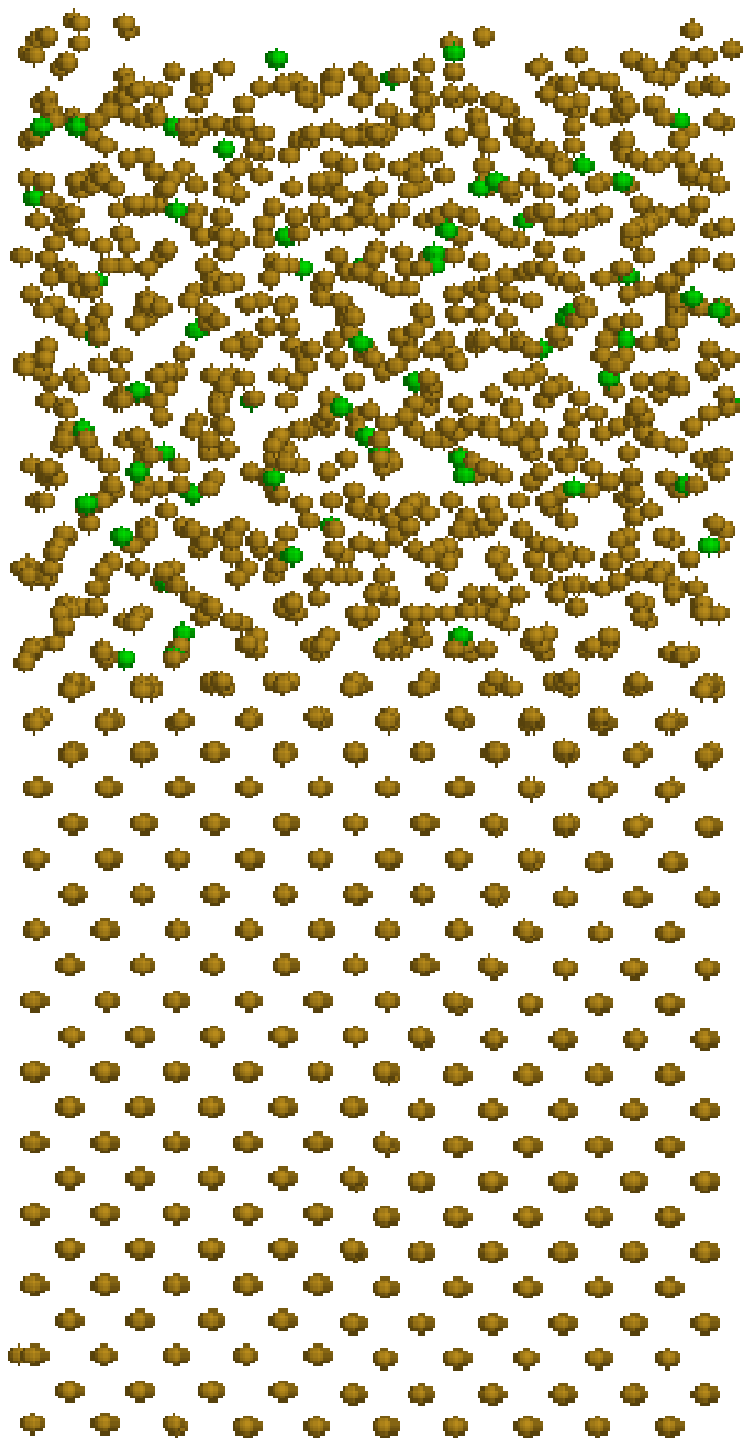}}
\resizebox{0.4\columnwidth}{!}{\includegraphics{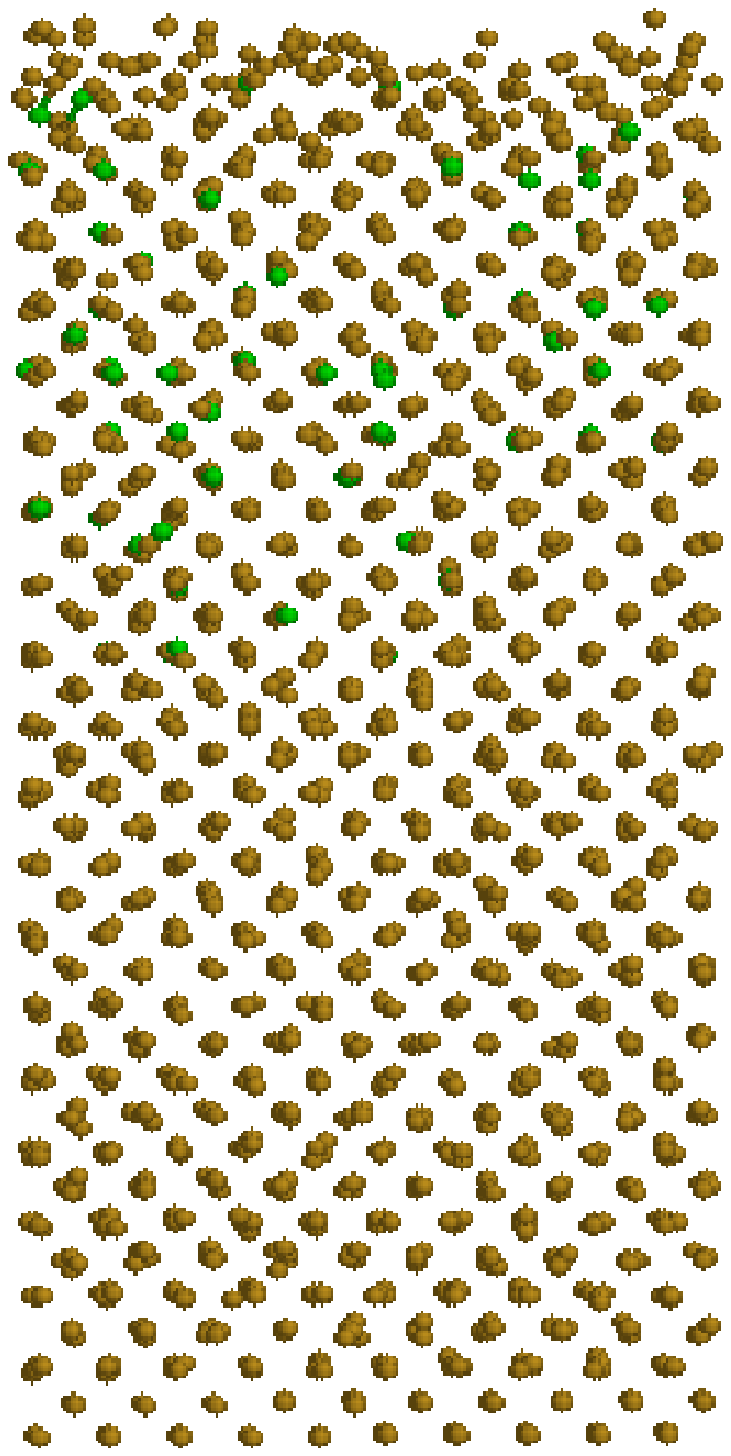}}
\caption{An example of an initial $a$-Si:B/$c$-Si $[100]$ interface for a 6 \% boron concentration (top panel). The next figure shows the configuration after annealing where the  dopant has been incorporated during re-growth. The segregation performed by the interface and the diffusion in the amorphous part can be clearly observed.}       
\label{fig:nine_figure}
\end{figure}

The possible influence of boron atoms in the silicon amorphous part has been investigated in the case of  interface $[100]$. The implantation step has not been  Fig. \ref{fig:nine_figure} presents an example of initial doped $a$-Si (doped)/$c$-Si interface  $[100]$ and illustrates the final state where  the amorphous part is completely recrystallised with most of the boron atoms incorporated  in substitutional positions of the crystalline matrix. This result is in agreement with a previous study on the annealing of damage generated by octodecaborane implantation \cite{Marques06}. The results concerning the interface recrystallisation velocity are first detailed just below and then discussed in further subsection.

\subsubsection{Interface recrystallisation velocity}
 
\begin{figure}[btp]
\centering
\resizebox{1\columnwidth}{!}{\includegraphics{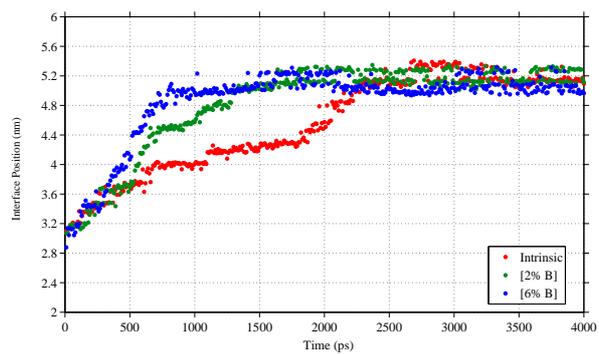}}
\caption{Example of interface migration for various dopant concentrations at 2000K as a function of the annealing time.}
\label{fig:ten_figure}
\end{figure}

\begin{figure}[btp]
\centering
\resizebox{1\columnwidth}{!}{\includegraphics{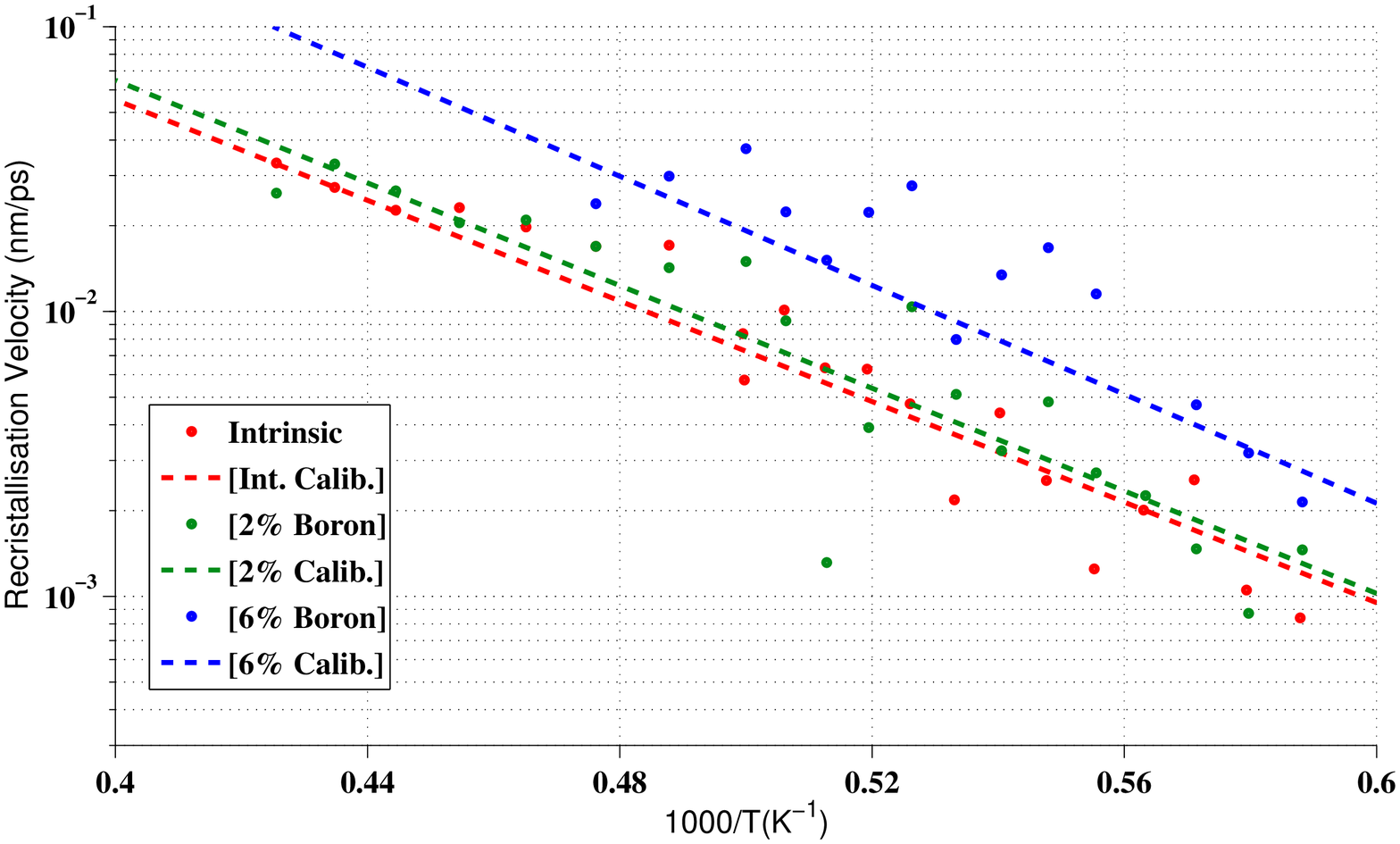}}
\caption{Recrystallisation velocity for solid phase epitaxy as a function of temperature and boron concentration in the initial amorphous phase.}
\label{fig:eleven_figure}
\end{figure}

Fig. \ref{fig:ten_figure} reports the evolution of the interface position at 2000K \footnote{which would correspond to a rescaled temperature of about 1000$^{\circ}$C} for different boron concentrations. In this case, it is clearly  shown that  the presence of boron is increasing the velocity of the interface migration. Fig. \ref{fig:eleven_figure} summarises all the computer experiments performed in the intrinsic case and for  2 \% and 6 \% boron concentration configurations. Basically, for most  temperatures, it can be observed that there is no significant difference between the intrinsic case and the 2 \% case.   The calibration by an Arrhenius law for the last case gives the same activation energy and a small difference for the prefactor (6.2 \%). For a very large boron concentration  (6 \%), a small increase in  the solid phase epitaxy velocity is observed leading to an increase in  the prefactor by a ratio 1.34.

\subsubsection{Discussion}

In this work, a complete different behaviour has been observed compared to similar  previous molecular dynamics simulations performed by Mattoni et al. \cite{Mattoni03}. Using molecular dynamics in the framework of the Stillinger-Weber empirical potential, an increase of the re-growth velocity   at low boron concentrations (below 1\%) has been observed whereas  for larger concentrations (1\% up to 10\%) the velocity is reduced.  On the experimental side  \cite{Olson88}, it has been observed that below 0.06 \%, the boron concentration has almost no influence on the recrystallisation velocity. However, around 0.2 \%, a significant increase is observed up to a factor 10 at a low temperature annealing ($\sim 600^{\circ}$C). Our results obtained using Tersoff are much more in agreement with the experimental picture and the observed minor influence of boron at low concentration. Probably, in the case of Stillinger-Weber simulation, the mechanism is more compatible with a framework of liquid phase epitaxy with the high tendency of SW based $a$-Si  for melting even at low temperature \cite{Krzeminski07}. Compared to the experimental results, it would not be so surprising that the influence of dopant is largely underestimated by our empirical molecular dynamics simulation since the  Fermi level  variations in our system are not taken into account theoretically. Generalised Fermi-Level Shifting (GFLS) model and  the influence of  charged defects at the interface are often assumed empirically with some success to estimate the enhancement of the normalised re-growth rate defined by the ratio of the doped recrystallisation velocity in the doped with the intrinsic one $\displaystyle \frac{v_{SPE}}{v_{int}}$ \cite{Johnson08} :

\begin{equation}
\displaystyle
\frac{v_{SPE}}{v_{int}}=\frac{1+g \exp \displaystyle \frac {(E_{f}-E_{d})}{kT}}{1+g \exp \displaystyle \frac{(E_{fi}-E_{d})}{kT}}
\end{equation}

where E$_{f}$  and E$_{d}$ are respectively the Fermi level and the defect level in the band gap whereas $g$ is the degeneracy factor. As pointed out by Johnson et al. \cite{Johnson08}  more work is needed in this direction in order to confirm the validity of the GFLS approach. First, a  more refined description level which incorporates  electronic structure and Fermi shift such as tight-binding  or ab-initio molecular dynamics would probably be mandatory for such a purpose. Next, the ideal system under study (doped $a$-Si:B/$c$-Si interface)  has probably to be improved in order to match the experimental picture  since for example dopant implantation is known to generate large amounts of defects  (end of range defects, interstitials ,...) and surface roughness.

				.

\section{Conclusions}

Empirical molecular dynamics simulation have been undertaken in order to investigate the silicon re-growth at the atomic level. Earlier results on the influence of the empirical potential and the large sensibility of the virtual silicon amorphous to the atomic force description have been again illustrated. On the other hand, it has been observed that the interface migration velocity estimated is not so sensitive to the way a virtual amorphous silicon cluster is generated provided the amorphous character is preserved during annealing. The influence of orientation effects has been investigated at the atomic level showing a large influence of the surface orientation  effects on the recrystallisation mechanism. The physical picture described by molecular dynamics is more complex than the classical re-growth theory since the relation between the atomistic structure of the interface and the defects generation is clearly underlined. The concept of a sharp interface often used in the classical theory must be put in perspective as molecular dynamics suggests a more diffuse interface. More attention is probably needed to investigate more  in depth the influence of defect on velocity re-growth. Concerning doping effects, a small increase in the recrystallisation velocity has been observed at large doping concentration. Further work is clearly needed in this direction to study  if the influence of charged  defects could be introduced empirically in molecular dynamics simulation in order to validate or not the generalised Fermi-level shifting model often assumed by the classical re-growth theory.

\begin{acknowledgement}
The authors are grateful to  Jean-Michel Droulez  for his regular support on information technology  aspects and to Fabrizio Cleri for stimulating discussion on molecular dynamics. This work has been supported partly by the European Commission as a contribution to the project PULLNANO under contract n$^{\circ}$ IST-026828 from the Information Society Technologies Programme (IST).
\end{acknowledgement}

\begin{table*}[tbp]
\centering
  \caption{Main  molecular dynamics parameters for disordered part generation.}
  \begin{tabular}{p{25mm}p{25mm}p{25mm}p{16mm}}
    \hline
    \hline
     Potential                  & Annealing Range (K)& Melting (K)& Quenching Rate (K/ps)\\
    \hline                             
     Tersoff                    & [1700-2500]       &  3500      &    $1 \times 10^{11}$\\
     SW (Standard)              & [1000-1700]       &  2200      &    $5 \times 10^{11}$\\
    \hline
  \end{tabular}
  \label{table:first_table}
\end{table*}

\clearpage

\begin{table*}[btp]
\centering
\caption{Distribution of the  mean coordination number (CN) \hspace{\linewidth} for the different initial disordered interfaces at 300K.}
\begin{tabular}{|c|c|c|c|c|c|c|c|c|}
\hline
Conditions&CN&2fold&3fold&4fold&5fold&6fold\\
\hline\hline
$a$-Si, WWW    &3.63&7.6 \%&19.7\%&70.8 \% &0.9\%  &0\%\\
$a$-Si, Tersoff&4.17&0.7 \%& 9.9\%&63.4 \% &24 \%  &1.9 \%\\
$l$-Si, SW     &2.92&24. \%&38.5\%&25.67\% &3.4 &0.16 \%\\
\hline\hline
\end{tabular}
\label{table:second_table}
\end{table*}

\clearpage

\clearpage

\clearpage

\clearpage

\clearpage

\clearpage

\clearpage

\clearpage

\clearpage

\clearpage

\clearpage
 
\begin{thebibliography}{}
\bibitem{Duffy07} R. Duffy, M. J. H.  Van Dal, B. J.  Pawlak, M.  Kaiser, R. G. R  Weemaes, B.  Degroote, E.  Kunnen, E. Altamirano, Appl. Phys. Lett., {\bf 90}, 241912, (2007).
\bibitem{Saenger07}    K. L. Saenger, J. P. de Souza, K. E. Fogel, J. A. Ott, C. Y. Sung, D. K. Sadana, H. Yin, J. Appl. Phys., {\bf 101}, 024908, (2007).
\bibitem{Agaiby}       R. M. B. Agaiby, M. Becker, S. B. Thapa, U. Urmoneit, A. Berger, A. Gawlik,  G. Sarau and S. H. Christiansen, J. App. Phys., {\bf 107}, 3319654, (2010).
\bibitem{Huet09}       S. Huet, P. Juliet, X. Ducros, F. Sanchette, United States Patent, 7592198 B2, (2009).
\bibitem{Duffy04}      R. Duffy, V. C. Venezia, A. Henriga, B. J. Pawlak, M. J. P. Hopstaken, G. C. J. Maas, Y. Tamminga, T. Dao, F. Roozeboom and L. Pelaz, Appl. Phys. Lett., {\bf 84}, 4283 (2004).
\bibitem{Chernov04}    A. A. Chernov, J. Cryst Growth, Vol. 264, Issue 4, p. 499-518, (2004).
\bibitem{BCF51}        W. K. Burton, N. Cabrera, F. C. Frank, Phil. Trans. R. Soc. Lond. A, Vol. 243, 866, p. 299-358, (1951).
\bibitem{Aziz91}      Guo-Quan Lu and Eric Nygren and Michael J. Aziz, J. Appl. Phys., Vol. 70, 5323, (1991).
\bibitem{Pelaz09}      L. Pelaz, L. A Marqu\'es, M. Aboy, P. L\'opez and I. Santos, J. Eur. Phys. B, {\bf 72}, 323 (2009). 
\bibitem{Cuny06}       V. Cuny, Q. Brulin, E. Lampin, E. Lecat, C. Krzeminski and F. Cleri, Europhys. Lett., {\bf 76}, 842, (2006).
\bibitem{Motooka00}    T. Motooka, K. Nisihira, S. Munetoh, K.  Moriguchi, A. Shintani, Phys. Rev. B, {\bf 61}, 8537, (2000).
\bibitem{Grabow89}    M. H. Grabow, G. H. Gilmer and A. F. Bakker, in Atomic Scale Calculations in Materials Science, edited by J. Tersoff, D. Vanderbilt, and V. Vitek, Mater. Res. Soc. Symp. Proc., Vol. {\bf 141}, Pittsburgh, (1989).

\bibitem{Bragado09}    I. Martin-Bragado and V. Moroz, Appl. Phys. Lett., {\bf 95}, 123123, (2009).
\bibitem{Ishimaru97}   M. Ishimaru, S. Munetoh, T. Motooka, Phys. Rev. B, {\bf 56}, 15133, (1997).
\bibitem{Nose-Hoover} S. Nos\'e, Molec. Phys.{\bf 52}, 255 (1984). W. G. Hoover, Phys. Rev. A {\bf 31}, 1695 (1985).
\bibitem{Tersoff}      J. Tersoff, Phys. Rev. B {\bf 38}, 9902 (1988).
\bibitem{SW}           F. H. Stillinger and T. A. Weber, Phys. Rev. B {\bf 31}, 5262 (1985).
\bibitem{Mattoni04}    A. Mattoni  and L. Colombo, Phys. Rev. B, {\bf 69}, 045204,  (2004).
\bibitem{Perez_Martin04} A. Mari Carmen P\'erez-Mart\'in, Jos\'e J. Jim\'enez-Rodriguez, Jos\'e Carlos Jim\'enez-S\'aez, Applied Surface Science, {\bf 234}, 228, (2004).
\bibitem{Krzeminski07} C. Krzeminski, Q. Brulin, V. Cuny, E. Lecat,  E. Lampin,  F. Cleri, J. Appl. Phys., {\bf 101}, 2743089 (2007).
\bibitem{Wooten85}     F. Wooten, K. Winer and D. Weaire, Phys. Rev. Lett. {\bf 54}, 1392 (1985).
\bibitem{Drosd82}      R. Drosd, J. Washburn, J. Appl. Phys., {\bf 53}, 1982.
\bibitem{Lampin09}     E. Lampin and C. Krzeminski, J. Appl. Phys., {\bf 106}, 063519 (2009).
\bibitem{Olson88}      G. L. Olson and J. A. Roth, Mater. Sci. Rep. {\bf 3}, 1 (1988).
\bibitem{Csepregi78}   Csepregi, E. F. Kennedy, J. W. Mayer, T. W. Sigmon, J. Appl. Phys.,{\bf 49}, (1978).

\bibitem{Posselt09}    M. Posselt and A. Gabriel, Phys. Rev. B, {\bf 80}, 045202, (2009).
\bibitem{Lu91}         G.-Q. Lu, E. Nygren and M. J Aziz, J. Appl. Phys., {\bf 70}, 5323, (1991).

\bibitem{Marques06}  L. A. Marqu\'es, L. Pelaz, I. Santos and V. C. Venezia, Phys. Rev. B, {\bf 74}, 201201(R), (2006).
 
\bibitem{Mattoni03}    A. Mattoni and L. Colombo, EPL (Europhysics Letters), {\bf 62}, 6, (2003).
\bibitem{Johnson08}    B. C. Johnson, P. Gortmaker, J. C. McCallum, Phys. Rev. B, {\bf 77}, 214109, (2008).
\end{thebibliography}
\end{document}